\documentclass[twocolumn]{aastex6}

\usepackage{bm}
\usepackage{amsmath}
\usepackage{xcolor}
\usepackage{graphicx}
\usepackage{hyperref}
\usepackage[all]{hypcap}

\newcommand{\suz}{\sp{(0)}}
\newcommand{\suo}{\sp{(1)}}
\newcommand{\sut}{\sp{(2)}}

\begin{document}

    \title{Propagation of waves in weakly ionized two-fluid plasmas. II. Nonlinear Alfvénic waves}
    \shorttitle{Nonlinear Alfvénic waves in weakly ionized plasmas}
    \shortauthors{Martínez-Gómez}
    
    \author{David Martínez-Gómez\altaffilmark{1,2}}
    \altaffiltext{1}{Departament de Física, Universitat de les Illes Balears, E-07122, Palma de Mallorca, Spain}
    \altaffiltext{2}{Institut d'Aplicacions Computacionals de Codi Comunitari (IAC3), Universitat de les Illes Balears, E-07122, Palma de Mallorca, Spain}
    \email{david.martinez@uib.es}
    
    \begin{abstract} 
        Weakly ionized plasmas can be found in the lower layers of the solar and stellar atmospheres and in structures such as prominences and spicules. A variety of density perturbations and bulk flows detected in these environments have been explained as the result of the ponderomotive force generated by nonlinear Alfvénic waves. In addition, the dissipation of the energy carried by these waves leads to heating of the plasma. Here, we use a two-fluid model to study the combined influence of Hall's current and elastic collisions between ions and neutrals on the propagation of linearly and circularly polarized transverse waves in weakly ionized plasmas. We derive analytical expressions for the damping and heating rates, showing their dependence on the strength of the collisional coupling and on the polarization state. We also perform numerical simulations to investigate the nonlinear generation of density perturbations and bulk flows related to the ponderomotive force and the energy dissipation by the ion-neutral interaction. We find that the nonlinear perturbations associated with the circularly polarized eigenmodes do not show the oscillatory motions typically caused by linearly polarized eigenmodes, but they retain the non-oscillatory bulk flows. We also briefly discuss how in weak coupling conditions the nonlinear dynamics of the neutral fluid is mainly driven by the wave energy dissipation while the ponderomotive force only directly acts on the charged fluid, resulting in different amplitudes of the longitudinal motions and the perturbations of density and temperature.
    \end{abstract}
    

\section{Introduction} \label{sec:intro}
    The study of the nonlinear evolution of transverse waves propagating parallel to the background magnetic field (which are collectively denoted as Alfvénic waves) is of special interest for understanding the dynamics of partially ionized plasmas. On the one hand, the collisional interaction between the electrically charged and the neutral components of the plasma leads to dissipation of the energy carried by the waves \citep[][]{Mestel1956MNRAS.116..503M,Piddington1956MNRAS.116..314P,Watanabe1961CaJPh..39.1044W} and to an increase of the temperature \citep[][]{Osterbrock1961ApJ...134..347O,Goodman2011ApJ...735...45G,Song2011JGRA..116.9104S,KhomenkoCollados2012ApJ...747...87K,Arber2016ApJ...817...94A,Soler2016A&A...592A..28S,Kuzma2020A&A...639A..45K}. On the other hand, this kind of waves may generate perturbations of density, pressure and the longitudinal component of the velocity \citep[][]{Hollweg1971JGR....76.5155H,Rankin1994JGR....9921291R,VERWICHTE_NAKARIAKOV_LONGBOTTOM_1999,Botha2000A&A...363.1186B,Zheng2016PhyS...91a5601Z}. These perturbations are the result of a ponderomotive force associated with the gradients in magnetic pressure. In addition, the dissipation of the wave energy causes the appearance of thermal pressure gradients that induce net motions in the longitudinal direction, usually referred to as bulk flows. In fully ionized plasmas the bulk flows can be caused by resistivity and viscosity \citep[][]{McLaughlin2011A&A...527A.149M,Threlfall2012PhDT.......253T,Zheng2016PhyS...91c5601Z}, while in partially ionized plasmas there is also a contribution from ion-neutral collisions \citep{MartinezGomez2018ApJ...856...16M,Ballester2020A&A...641A..48B,Ballester2024RSPTA.38230222B}.

    In the context of solar and stellar atmospheres, the ponderomotive force has been invoked as an explanation for the first ionization potential (FIP) effect \citep{Laming2004ApJ...614.1063L,Laming2015LRSP...12....2L,To2026arXiv260413174T}. This effect refers to the observed enhancement in the coronal abundances of elements with a first ionization potential below $10 \ \rm{eV}$ (known as low-FIP elements) in comparison to their abundances in the lower layers of the atmosphere, while the abundances of high-FIP elements are not affected \citep{Testa2023ApJ...944..117T,Murabito2024PhRvL.132u5201M}. The ponderomotive force directly acts on the ionized species but not on the neutral ones and in the lower layers of the solar and stellar atmospheres, low-FIP elements tend to be ionized while high-FIP elements remain neutral, so the former are more efficiently pushed upwards \citep{Testa2010SSRv..157...37T,Testa2015RSPTA.37340259T,MartinezSykora2023ApJ...949..112M}.

    The nonlinear dynamics of transverse waves may also explain density enhancements at the tops of coronal loops, in which large-amplitude standing kink oscillations give rise to plasma upflows from their legs \citep{Terradas2004ApJ...610..523T,Terradas2022A&A...660A..24T}. Moreover, the ponderomotive coupling between propagating transverse waves and longitudinal motions has been used to describe the formation of solar spicules \citep{Haerendel1992Natur.360..241H,DePontieu1998A&A...338..729D,Kudoh1999ApJ...514..493K,Brady2016ApJ...829...80B,MartinezSykora2017Sci...356.1269M} and the generation of plasma outflows in the solar chromosphere \citep[see, e.g.,][]{Pelekhata2021A&A...652A.114P,Pelekhata2023A&A...669A..47P,Kumar2024A&A...681A..60K}. In these scenarios, the energy dissipation by ion-neutral collisions enhances the plasma upflows \citep{MartinezSykora2023ApJ...949..112M,Srivastava2024RSPTA.38230220S}.

    In addition, it has been shown that the ion-neutral interaction modifies the spatial and temporal scales in which Hall's current strongly impacts the plasma dynamics \citep{Amagishi1993PhRvL..71..360A,Pandey2008MNRAS.385.2269P,Pandey2015MNRAS.447.3604P,GonzalezMorales2020A&A...642A.220G,Khomenko2021RSPTA.37900176K}. One of its main effects is inducing a rotation in the oscillation plane of waves, so they become circularly polarized and their behavior strongly depends on their polarization state \citep{Lighthill1960RSPTA.252..397L,Stix1992wapl.book.....S}. For instance, the ion-cyclotron waves, with left-handed polarization, are affected by resonances and cut-off regions associated with the cyclotron frequencies and propagate with a smaller speed than the whistler waves, which have right-handed polarization \citep[see, e.g.,][]{Cramer2001paw..book.....C,Huba2003LNP...615..166H}. Moreover, these waves are dispersive and do not fulfill the classical energy equipartition relation \citep{Walen1944ArMAF..30A...1W,Ferraro1958ApJ...127..459F} but the kinetic energy dominates over the magnetic energy for the case of the ion-cyclotron modes while the opposite occurs for the whistler modes \citep{Campos1992_10.1063/1.860136}.

    The study of the influence of Hall's current on waves propagating in weakly ionized plasmas has typically focused on their linear stage \citep[see, e.g.,][]{Pandey2018MNRAS.476..344P,Pandey2022MNRAS.513.1842P,MartinezGomez2025ApJ...982....4M}. Therefore, in the present work, which is a direct continuation of \citet[][hereafter, Paper I]{MartinezGomez2025ApJ...982....4M}, we are interested in analyzing the nonlinear behavior of Alfvénic waves. To that goal we use a two-fluid plasma model \citep[][]{Mouschovias2011MNRAS.415.1751M,Soler2013ApJS..209...16S,Soler2013ApJ...767..171S,Khomenko2014PhPl...21i2901K} which allows us to explore a wide range of scenarios regarding the ion-neutral collisional coupling. By means of analytical computations and numerical simulations, we investigate the properties of small-amplitude transverse waves propagating along the parallel direction to the background magnetic field in a homogeneous medium. We consider the cases of linearly polarized waves without the influence of Hall's current (which, for the sake of simplicity, will be denoted as Alfvén modes) and of circularly polarized ion-cyclotron and whistler waves. Then, we describe how the wave damping, heating rates, wave energy distribution, and the evolution of densities, temperatures, and longitudinal bulk flows depend on the polarization state of the waves and on the degree of collisional coupling.
    
\section{Model and methods} \label{sec:model_methods}
\subsection{Two-fluid plasma model and background atmosphere} \label{sec:2F_model}
    We assume that there is a strong collisional coupling between electrons, denoted by the subscript `e', and all the ionized species of the plasma, denoted by `i', so they can be treated together as a single charged fluid, denoted by `c'. Likewise, all neutral species are combined into a single neutral fluid, denoted by `n'. Thus, we use a two-fluid approach \citep[see, e.g.,][]{Mouschovias2011MNRAS.415.1751M,Zaqarashvili2011A&A...529A..82Z,Soler2013ApJ...767..171S,Khomenko2014PhPl...21i2901K} in which the charged and neutral fluids interact through elastic collisions. This interaction is described by including momentum and energy transfer terms  \citep{Schunk1977RvGSP..15..429S,Draine1986MNRAS.220..133D} in the corresponding evolution equations. Other mechanisms related to the partial ionization of the plasma, such as charge-exchange collisions, ionization or recombination \citep[see, e.g.,][]{Meier2012PhPl...19g2508M,Leake2012ApJ...760..109L,Popescu2019A&A...627A..25P,Snow2021A&A...645A..81S} are not taken into account. Moreover, we consider the influence of Hall's current \citep{Lighthill1960RSPTA.252..397L,Cramer2001paw..book.....C} but not that of other non-ideal processes such as ohmic diffusion, viscosity or thermal conduction. We also assume that the pressure ($P_{\rm{s}}$), number density ($n_{\rm{s}}$) and temperature ($T_{\rm{s}}$) of each fluid `s' are related by the ideal equation of state. Therefore, the full nonlinear equations that describe the plasma dynamics are given by Eqs. (1) - (12) of \hyperlink{PaperI}{Paper I}.

    As background atmosphere we consider an infinite homogeneous medium embedded in a uniform magnetic field of strength $B_{\rm{0}}$. We do not take into account effects related to the stratification caused by the presence of gravity, such as the reflection of waves \citep[see, e.g.,][]{Ferraro1954ApJ...119..393F,Pinto2012A&A...544A..66P,Soler2026A&A...708A..68S} or the variation of their amplitudes as they propagate through the atmosphere \cite[see, e.g.,][]{Campos1992_10.1063/1.860136,Zaqarashvili2013A&A...549A.113Z,Popescu2019A&A...627A..25P,Popescu2019A&A...630A..79P,Kuzma2020A&A...639A..45K}.

\subsection{Regular perturbations approach} \label{sec:expansion}
    In order to separate the linear behavior of the waves from their nonlinear dynamics, we first perform a regular perturbations expansion \citep[see, e.g.,][]{Bender1978amms.book.....B}. We rewrite every variable in the two-fluid equations in the form
    \begin{equation}
        \bm{f} = \sum_{n=0}^{n_{\rm{max}}} \epsilon^{n}\bm{f}^{(n)},
    \end{equation}
    where the term $\bm{f}^{(0)}$ represents the background value, the terms $\bm{f}^{(n)}$ represent the perturbations of order $n$ (with $n = 1$ corresponding to the linear regime, and $n \geq 2$ corresponding to the nonlinear regime), $n_{\rm{max}}$ is the maximum order to be considered in the expansion, and $\epsilon$ is a small parameter related to one of the characteristic speeds of the system \citep[see, e.g.,][]{Rankin1994JGR....9921291R, Ballester2020A&A...641A..48B}. For the present study, the reference speed will be the equilibrium Alfvén speed (denoted by $c_{\rm{A}}$), so the parameter $\epsilon$ will be defined as $\epsilon = V_{\rm{0}} / c_{\rm{A}}$, where $V_{\rm{0}}$ is the initial amplitude of the velocity perturbations. Then, assuming that $\epsilon \ll 1$, the different terms of the evolution equations are gathered according to their powers of $\epsilon$ to obtain separate systems of equations for each order of the expansion.

    In this work we consider a static uniform medium, so the zeroth-order terms for the density ($\rho_{\rm{s}}\suz \equiv n_{\rm{s}}\suz m_{\rm{s}}$, where $m_{\rm{s}}$ is the mass of a particle of the fluid), pressure ($P_{\rm{s}}\suz$) and temperature ($T_{\rm{s}}\suz$) of each fluid `s' are constant, and the are no background velocities, $\bm{V}_{\rm{s}}\suz = 0$. In addition, we assume that the background magnetic field is oriented along the $z$-direction, so $\bm{B}\suz = \left(0, 0, B_{\rm{0}} \right)$.

    Then, we analyze the properties of perturbations of up to the second order of the expansion, focusing on the propagation along the parallel direction to the background magnetic field. Therefore, we impose that the derivatives with respect to the $x$ and $y$ directions are zero.

\subsection{Numerical simulations} \label{sec:methods_sims}
    To complement the analytical results, we perform 1D numerical simulations with a multi-fluid version \citep{MartinezGomez2016ApJ...832..101M,MartinezGomez2017ApJ...837...80M} of the MoLMHD code \citep{Terradas2008ApJ...675..875T,Terradas2013ApJ...778...49T}. This code uses the method of lines \citep{SARMIN19631537,schiesser1991numerical} and in this version the spatial derivatives are computed through a central finite differences scheme of fourth order of accuracy while the the temporal evolution is computed by means of an explicit third-order Runge-Kutta method \citep{Harten1983JCoPh..49..357H}. The code is parallelized through OpenMP routines and uses artificial dissipation terms of fourth order to ensure the stability of the simulations \citep{Kreiss1973}.

    We simulate the propagation of waves generated by a periodic driver by applying at $z = 0$ appropriate sinusoidal functions with a fixed frequency $\omega$ to the variables of the multi-fluid system. The specific details for each simulation are provided later (see Eqs. (\ref{eq:1order_alfven}) and (\ref{eq:1order_circularly})). At the opposite edge of the numerical domain, $z = z_{\rm{max}}$, we impose extrapolated boundary conditions. Moreover, in order to reduce the influence of transient effects during the first stages of the simulations, we multiply the driver by a smoothing function,
    \begin{equation} \label{eq:smooth}
        r(t) = \frac{1}{2}\left[1 + \tanh \left(\frac{t-N\tau}{W \tau} \right) \right],
    \end{equation}
    where $\tau = 2 \pi / \omega$ is the period of the wave, and the parameters $N$ and $W$ control the duration and smoothness of this initial phase of the simulations. For the cases analyzed here, we set $N = 1.5$ and $W = 0.5$, so the driver reaches its maximum amplitude after approximately three wave periods.
    
\section{Results for first-order perturbations} \label{sec:first_order}

\subsection{First-order equations} \label{sec:first_order_eqs}
    Gathering the terms proportional to $\epsilon$ from the expansion described in Section \ref{sec:expansion}, we obtain the system of equations that rules the temporal evolution of the linear perturbations propagating along $\bm{B}\suz$. 
    
    In order to simplify the notation, we use the following definitions: $\bm{V}_{\rm{s}\perp} \equiv \left(V_{\rm{s}x}\suo, V_{\rm{s}y}\suo, 0 \right)$ and $\rho_{\rm{s0}} \equiv \rho_{\rm{s}}\suz$ for $s \in \{c, n\}$, $\bm{B}_{\perp} \equiv \left(B_{x}\suo, B_{y}\suo, 0 \right)$, and $n_{\rm{e0}} \equiv n_{\rm{e}}\suz$. Thus, the equations for the first-order transverse perturbations are given by:
    \begin{equation} \label{eq:vn1_perp}
        \rho_{\rm{n0}} \frac{\partial \bm{V}_{\rm{n}\perp}}{\partial t} = \alpha_{\rm{cn}}\suz \left(\bm{V}_{\rm{c}\perp} - \bm{V}_{\rm{n}\perp} \right),
    \end{equation}

    \begin{equation} \label{eq:vc1_perp}
        \rho_{\rm{c0}} \frac{\partial \bm{V}_{\rm{c}\perp}}{\partial t} = \frac{B_{\rm{0}}}{\mu_{\rm{0}}}\frac{\partial \bm{B}_{\perp}}{\partial z} + \alpha_{\rm{cn}}\suz \left(\bm{V}_{\rm{n}\perp} - \bm{V}_{\rm{c}\perp} \right),
    \end{equation}

    \begin{equation} \label{eq:bx1}
        \frac{\partial B_{x}\suo}{\partial t} = B_{\rm{0}} \frac{\partial V_{\rm{c}x}\suo}{\partial z} + \frac{B_{\rm{0}}}{e n_{\rm{e0}} \mu_{\rm{0}}} \frac{\partial^{2} B_{y}\suo}{\partial z^{2}},
    \end{equation}
    and
    \begin{equation} \label{eq:by1}
        \frac{\partial B_{y}\suo}{\partial t} = B_{\rm{0}} \frac{\partial V_{\rm{c}y}\suo}{\partial z} - \frac{B_{\rm{0}}}{e n_{\rm{e0}}\mu_{\rm{0}}} \frac{\partial^{2} B_{x}\suo}{\partial z^{2}},
    \end{equation}
    where $\alpha_{\rm{cn}}\suz$ is the background value of the charge-neutral friction coefficient \citep[see, e.g.,][]{Braginskii1965RvPP....1..205B,Draine1986MNRAS.220..133D}, $\mu_{\rm{0}}$ is the vacuum magnetic permeability, and $e$ is the elementary electric charge. These equations show that the transverse perturbations are not coupled to the longitudinal or to the density and pressure perturbations in the first-order level. Thus, we can assume that $V_{\rm{s}z}\suo = \rho_{\rm{s}}\suo = P_{\rm{s}}\suo = 0$ for $s \in \{c, n\}$ and $B_{z}\suo = 0$.

\subsection{Wavenumber and quality factor} \label{sec:kz_qfactor}
    As shown in \hyperlink{PaperI}{Paper I}, if we assume that the linear perturbations are proportional to $\exp(i k_{z}z - i \omega t)$, where $k_{z}$ is the longitudinal wavenumber, the following dispersion relation can be obtained from Eqs. (\ref{eq:vn1_perp})-(\ref{eq:by1}):
    \begin{equation} \label{eq:kz2_hall_gamma}
        k_{z}^{2} = \frac{\omega^{2}}{c_{\rm{A}}^{2}}\frac{\omega + i \left(1 + \chi \right) \nu_{\rm{nc}}}{\omega \Gamma_{\rm{i}} + i \nu_{\rm{nc}} \Gamma_{\rm{H}}},
        \end{equation}
    where $c_{\rm{A}} = B_{\rm{0}} / \sqrt{\mu_{\rm{0}} \rho_{\rm{c0}}}$ is the Alfvén speed of the charged fluid, $\nu_{\rm{nc}} = \alpha_{\rm{cn}}\suz / \rho_{\rm{n0}}$ is the neutral-charge collision frequency, $\chi = \rho_{\rm{n0}}/ \rho_{\rm{c0}}$ is the neutral-to-charge density ratio, and
    \begin{equation} \label{eq:gamma_delta}
        \Gamma_{\rm{i}} = 1 \mp \frac{\omega}{\Omega_{\rm{i}}} \quad \text{and} \quad \Gamma_{\rm{H}} = 1 \mp \frac{\omega}{\Omega_{\rm{H}}}
    \end{equation}
    are real parameters associated with the ion cyclotron frequency, $\Omega_{\rm{i}} = e B_{\rm{0}} / m_{\rm{c}}$, and the Hall frequency \citep{Pandey2008MNRAS.385.2269P,Pandey2015MNRAS.447.3604P}, $\Omega_{\rm{H}} = \Omega_{\rm{i}}/\left(1 + \chi \right)$, respectively. The ``$-$" sign corresponds to ion-cyclotron waves while the ``$+$" corresponds to whistler waves.

    Exact solutions of the dispersion relation can be obtained by splitting the wavenumber in its real and imaginary parts \citep[see, e.g.,][]{Soler2013ApJ...767..171S}, $k_{z} = k_{\rm{R}} + i k_{\rm{I}}$, substituting this expression into Equation (\ref{eq:kz2_hall_gamma}) and solving the corresponding system of equations for $k_{\rm{R}}$ and $k_{\rm{I}}$. The solutions are given by:
    \begin{eqnarray} 
        k_{\rm{R}}^{2} = \frac{1}{2}\frac{\omega^{2}}{c_{\rm{A}}^{2}}\Bigg[\frac{\omega^{2} \Gamma_{\rm{i}} + \left(1 + \chi \right) \nu_{\rm{nc}}^{2} \Gamma_{\rm{H}}}{\omega^{2} \Gamma_{\rm{i}}^{2} + \nu_{\rm{nc}}^{2} \Gamma_{\rm{H}}^{2}} \nonumber \\
        + \left(\frac{\omega^{2} + \left(1 + \chi \right)^{2} \nu_{\rm{nc}}^{2}}{\omega^{2} \Gamma_{\rm{i}}^{2} + \nu_{\rm{nc}}^{2} \Gamma_{\rm{H}}^{2}}\right)^{1/2} \Bigg]
        \label{eq:kR2}
    \end{eqnarray}
    and
    \begin{equation} \label{eq:kI2}
        k_{\rm{I}}^{2} = k_{\rm{R}}^{2} - \frac{\omega^{2}}{c_{\rm{A}}^{2}} \frac{\omega^{2} \Gamma_{\rm{i}} + \left(1 + \chi \right) \nu_{\rm{nc}}^{2} \Gamma_{\rm{H}}}{\omega^{2} \Gamma_{\rm{i}}^{2} + \nu_{\rm{nc}}^{2} \Gamma_{\rm{H}}^{2}}.
    \end{equation}

    In the limit of low frequencies, with $\omega \ll \{\Omega_{\rm{i}}, \Omega_{\rm{H}} \}$ so $\{\Gamma_{\rm{i}}, \Gamma_{\rm{H}} \} \to 1$, we recover Eqs. (35) and (36) from \citet{Soler2013ApJ...767..171S} for Alfvén waves when Hall's term is not taken into account.
    
    The quality factor of the waves, commonly defined as
    \begin{equation} \label{eq:qfactor}
        Q = \frac{1}{2}\frac{|k_{\rm{R}}|}{|k_{\rm{I}}|},
    \end{equation}
    provides a useful measure of the relevance of the damping rate in comparison with the oscillation rate. Combining Eqs. (\ref{eq:kR2}) - (\ref{eq:qfactor}), it is possible to obtain the following exact analytical expression:
    \begin{gather}
        Q = \frac{\omega^{2} \Gamma_{\rm{i}} + \left(1 + \chi \right) \nu_{\rm{nc}}^{2} \Gamma_{\rm{H}}}{2\chi \nu_{\rm{nc}} \omega} \nonumber \\
        + \frac{\Big[\left(\omega^{2} + \left(1 + \chi \right)^{2} \nu_{\rm{nc}}^{2} \right) \Big(\omega^{2} \Gamma_{\rm{i}}^{2} + \nu_{\rm{nc}}^{2} \Gamma_{\rm{H}}^{2} \Big)\Big]^{1/2}}{2\chi \nu_{\rm{nc}} \omega}.
        \label{eq:qfactor_exact}
    \end{gather}

    In the weak coupling regime ($\omega \gg \nu_{\rm{nc}}$), Eq. (\ref{eq:qfactor_exact}) reduces to
    \begin{equation} \label{eq:qfactor_weak}
        Q_{\rm{W}} \approx \frac{\omega \Gamma_{\rm{i}}}{\nu_{\rm{cn}}},
    \end{equation}
    where $\nu_{\rm{cn}} = \chi \nu_{\rm{nc}}$, while in the strong coupling regime ($\omega \ll \nu_{\rm{nc}}$), it reduces to
    \begin{equation} \label{eq:qfactor_strong}
        Q_{\rm{S}} \approx \frac{\left(1 + \chi \right) \nu_{\rm{nc}} \Gamma_{\rm{H}}}{\chi \omega}.
    \end{equation}

    Taking into account that $\{\Gamma_{\rm{i}}, \Gamma_{\rm{H}} \} \approx 1$ for low frequency Alfvén waves, while $\{\Gamma_{\rm{i}}, \Gamma_{\rm{H}} \} < 1$ for ion-cyclotron waves but $\{\Gamma_{\rm{i}}, \Gamma_{\rm{H}}\} > 1$ for whistler waves, we find that ion-cyclotron waves have the smallest quality factor of the three type of waves (resulting in a stronger damping), while the whistler modes are the least affected by the collisional damping.

\subsection{Eigenfunction relations} \label{sec:eigenfunction}
    From Eqs. (\ref{eq:vn1_perp}) and (\ref{eq:vc1_perp}), it is possible to obtain the following relations for the transverse perturbations of velocities and magnetic field \citep[see, e.g.,][]{Pinto2012A&A...544A..66P}:
    \begin{equation} \label{eq:vnvc}
        \bm{V}_{\rm{n\perp}} = \frac{i \nu_{\rm{nc}}}{\omega + i \nu_{\rm{nc}}}\bm{V}_{\rm{c\perp}} = \tilde{R}_{\rm{V}} \bm{V}_{\rm{c}\perp}
    \end{equation}
    and
    \begin{equation} \label{eq:b1vc}
        \bm{B}_{\perp} = - \Big(\frac{\omega^{2} + i \left(1 + \chi \right) \nu_{\rm{nc}} \omega}{\omega + i \nu_{\rm{nc}}}\Big) \Big(\frac{\mu_{\rm{0}} \rho_{\rm{c0}}}{k_{\rm{z}} B_{\rm{0}}}\Big) \bm{V}_{\rm{c\perp}} = \tilde{R}_{\rm{B}} \bm{V}_{\rm{c}\perp},
    \end{equation}
    where $\tilde{R}_{\rm{V}}$ and $\tilde{R}_{\rm{B}}$ are the ratios between the perturbations.

    For variables of the form $\bm{f} = \left(f_{x}, f_{y}, 0 \right)$, linearly polarized, left-handed circularly polarized and right-handed circularly polarized waves propagating along the $z$-direction fulfill \citep[][]{Cramer2001paw..book.....C,Bittencourt2004fopp.book.....B}
    \begin{equation} \label{eq:fyfx_rels}
        f_{x} = a f_{y}, \quad f_{x} = i f_{y}, \quad \text{and} \quad f_{x} = -i f_{y}, 
    \end{equation}
    respectively, where $a$ has real values. Then, assuming that $\bm{f} = \bm{\widetilde{f}} \exp\left(i k_{z} z - i \omega t \right)$, where $\bm{\widetilde{f}}$ is a complex quantity and $k_{z} = k_{\rm{R}} + i k_{\rm{I}}$, the real first-order perturbations for propagating waves can be written in the general form $\bm{f} = |\bm{\widetilde{f}}| \exp\left(-k_{\rm{I}}z\right) \sin \left(k_{\rm{R}} z - \omega t + \phi \right)$, where $\phi$ represents a phase shift with respect to a chosen reference perturbation. Here, we choose the $y$-component of the velocity of charges as the reference perturbation, expressed as $V_{cy}\suo = A_{\rm{c}} \exp \left(-k_{\rm{I}}z \right) \sin \left(k_{\rm{R}}z - \omega t \right)$, where $A_{\rm{c}}$ is a real amplitude. Therefore, in the remainder of this work, the following cases for the first-order perturbations are considered:

    \textit{1) Linear polarization:}
    \begin{subequations} \label{eq:1order_alfven}
        \begin{equation} \label{eq:vcy_alfven}
            V_{\rm{c}x}\suo = V_{\rm{c}y}\suo = A_{\rm{c}} \exp \left(-k_{\rm{I}}z \right) \sin \left(k_{\rm{R}}z - \omega t \right),
        \end{equation}
        \begin{equation} \label{eq:vny_alfven}
            V_{\rm{n}x}\suo = V_{\rm{n}y}\suo = A_{\rm{n}} \exp \left(-k_{\rm{I}}z \right) \sin \left(k_{\rm{R}}z - \omega t + \phi_{\rm{n}}\right),
        \end{equation}
        \begin{equation} \label{eq:by_alfven}
            B_{x}\suo = B_{y}\suo = A_{\rm{B}} \exp \left(-k_{\rm{I}}z \right) \sin \left(k_{\rm{R}}z - \omega t + \phi_{\rm{B}}\right),
        \end{equation}
    \end{subequations}
    where we have set $a = 1$ in the relations given by Eq. (\ref{eq:fyfx_rels});

    \textit{2) Circular polarization:}
    \begin{subequations} \label{eq:1order_circularly}
        \begin{equation} \label{eq:vcx_circ}
            V_{\rm{c}x}\suo = A_{\rm{c}} \exp \left(-k_{\rm{I}}z \right) \cos \left(k_{\rm{R}}z - \omega t \right),
        \end{equation}
        \begin{equation} \label{eq:vcy_circ}
            V_{\rm{c}y}\suo = \pm A_{\rm{c}} \exp \left(-k_{\rm{I}}z \right) \sin \left(k_{\rm{R}}z - \omega t \right),
        \end{equation}
        \begin{equation} \label{eq:vnx_circ}
            V_{\rm{n}x}\suo = A_{\rm{n}} \exp \left(-k_{\rm{I}}z \right) \cos \left(k_{\rm{R}}z - \omega t + \phi_{\rm{n}}\right),
        \end{equation}
        \begin{equation} \label{eq:vny_circ}
            V_{\rm{n}y}\suo = \pm A_{\rm{n}} \exp \left(-k_{\rm{I}}z \right) \sin \left(k_{\rm{R}}z - \omega t + \phi_{\rm{n}}\right),
        \end{equation}
        \begin{equation} \label{eq:bx_circ}
            B_{x}\suo = A_{\rm{B}} \exp \left(-k_{\rm{I}}z \right) \cos \left(k_{\rm{R}}z - \omega t + \phi_{\rm{B}}\right),
        \end{equation}
        \begin{equation} \label{eq:by_circ}
            B_{y}\suo = \pm A_{\rm{B}} \exp \left(-k_{\rm{I}}z \right) \sin \left(k_{\rm{R}}z - \omega t + \phi_{\rm{B}}\right),
        \end{equation}
    \end{subequations}
    where the ``$+$'' sign corresponds to the left-handed ion-cyclotron modes and the ``$-$'' sign corresponds to the right-handed whistler modes.

    \begin{figure*}
        \includegraphics[width=0.32\hsize]{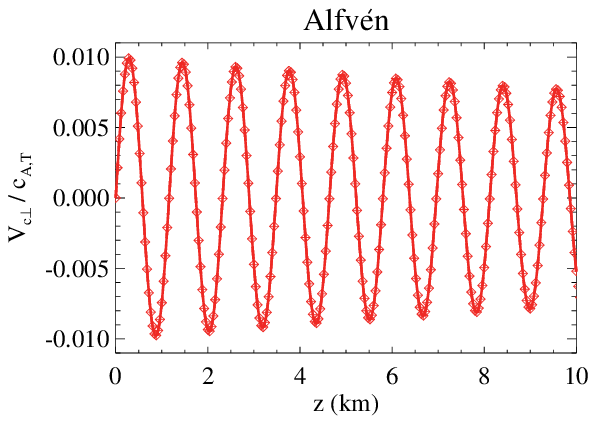}
        \includegraphics[width=0.32\hsize]{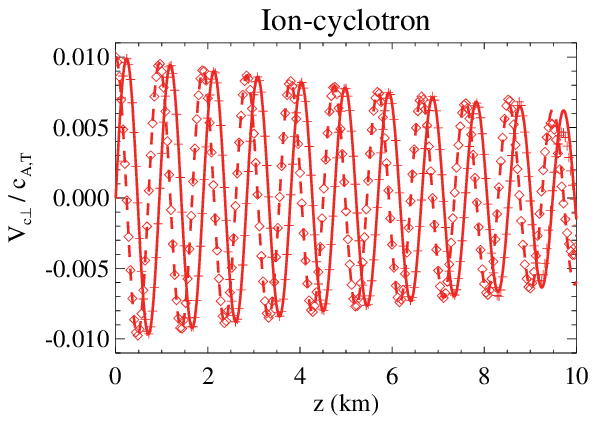}
        \includegraphics[width=0.32\hsize]{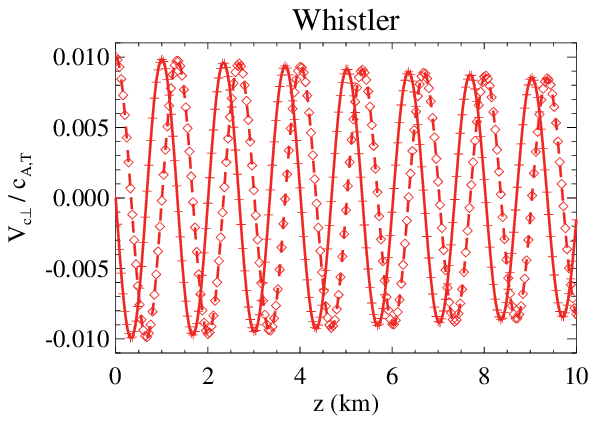} \\
        \includegraphics[width=0.32\hsize]{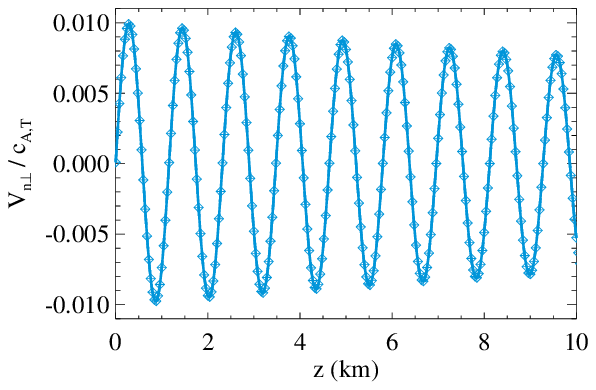}
        \includegraphics[width=0.32\hsize]{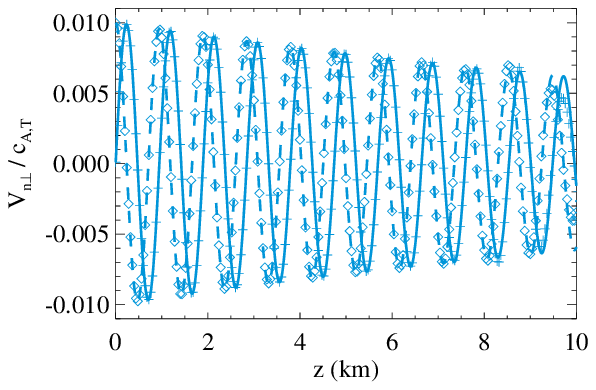}
        \includegraphics[width=0.32\hsize]{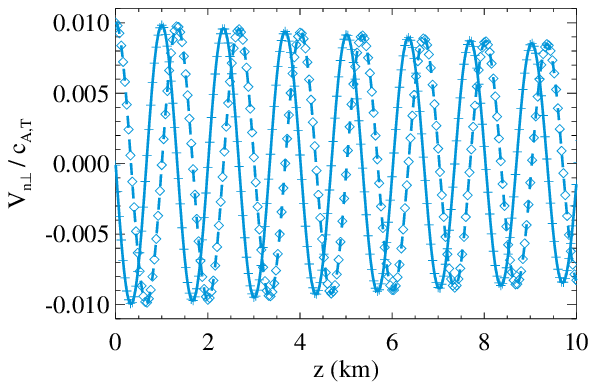} \\
        \includegraphics[width=0.32\hsize]{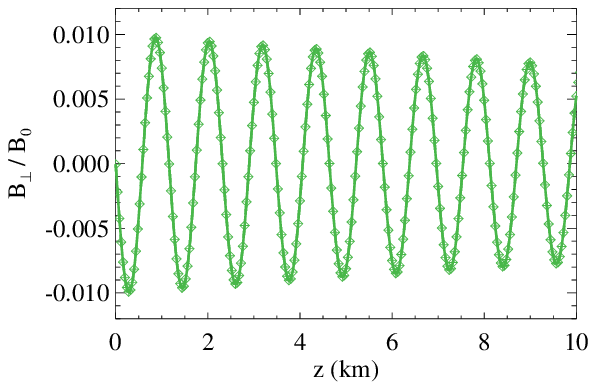}
        \includegraphics[width=0.32\hsize]{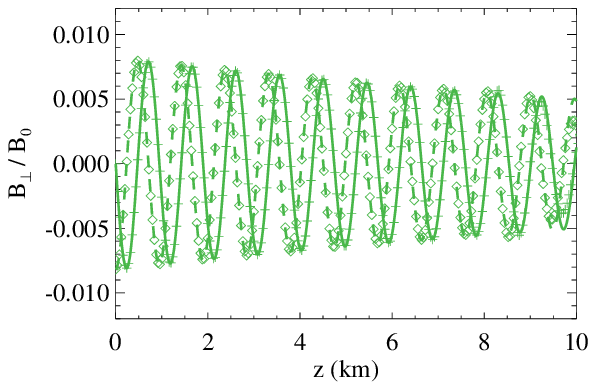}
        \includegraphics[width=0.32\hsize]{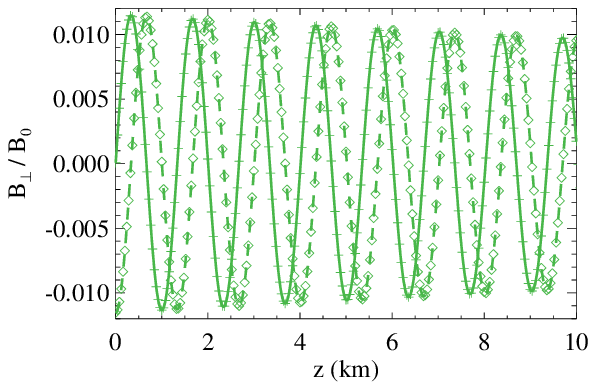}
        \caption{Normalized transverse perturbations of velocity of charges (top row), velocity of neutrals (middle row) and magnetic field (bottom row) at the time $t = 15 \tau$ for Alfvén waves (left column), ion-cyclotron waves (centre column), and whistler waves (right column) with a period of $\tau = 0.02$ and amplitude $A_{\rm{c}} = 0.01 \ \rm{c_{\rm{A,T}}}$. Case of a plasma with $n_{\rm{p}}=1.4 \times 10^{15} \ \rm{m^{-3}}$, $\chi = 100$, $B_{\rm{0}} = 10 \ \rm{G}$, and $\nu_{\rm{nc}} = 100 \omega$. Symbols represent the results from numerical simulations (diamonds: $x$-component; crosses: $y$-component), while lines represent the analytical expressions given by Eqs. (\ref{eq:1order_alfven}) and (\ref{eq:1order_circularly}).}
        \label{fig:firstorder_strong}
    \end{figure*}

    \begin{figure}
        \centering
        \includegraphics[width=0.9\hsize]{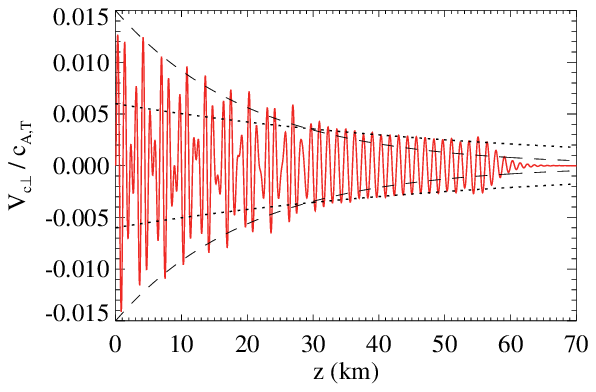} \\
        \includegraphics[width=0.9\hsize]{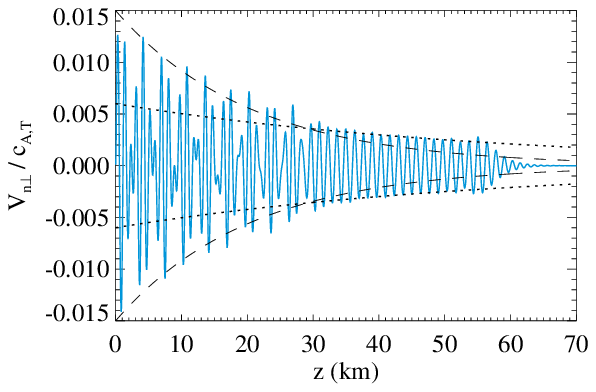} \\
        \includegraphics[width=0.9\hsize]{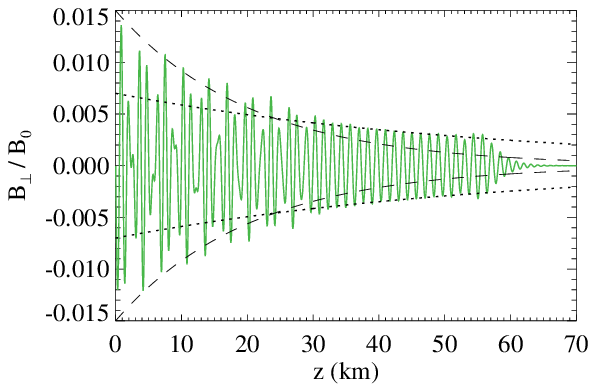}
        \caption{Normalized transverse perturbations of velocity of charges (top row), velocity of neutrals (middle row) and magnetic field (bottom row) at the time $t = 40 \tau$ from simulations of waves generated by the linearly polarized driver given by Eqs. (\ref{eq:1order_alfven}) but taking into account the effect of Hall's term. Same plasma parameters as in Figure \ref{fig:firstorder_strong} have been used. The black dashed and dotted lines represent the damping associated to the ion-cyclotron and the whistler modes, respectively. Only the $y$-component of the perturbations is shown.}
        \label{fig:firstorder_mix}
    \end{figure}

    \begin{figure*}
        \includegraphics[width=0.32\hsize]{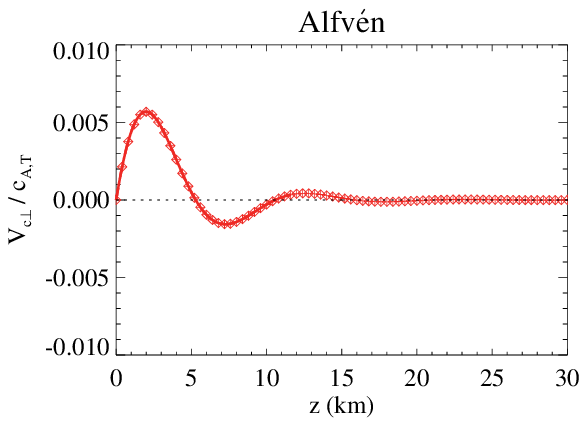}
        \includegraphics[width=0.32\hsize]{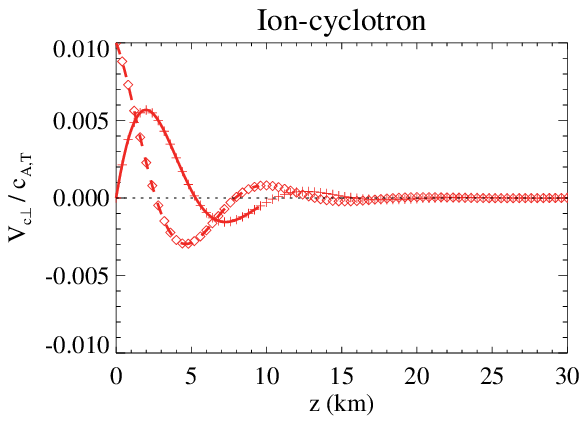}
        \includegraphics[width=0.32\hsize]{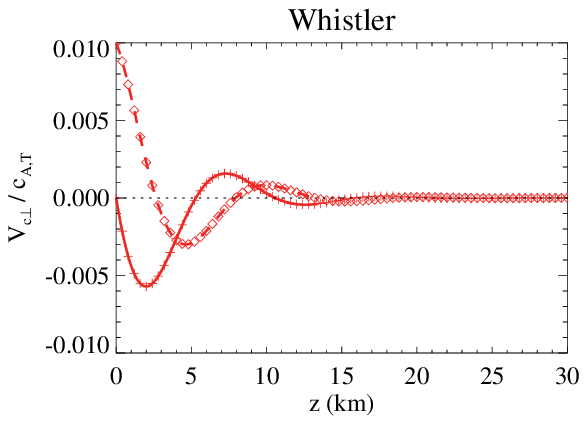} \\
        \includegraphics[width=0.32\hsize]{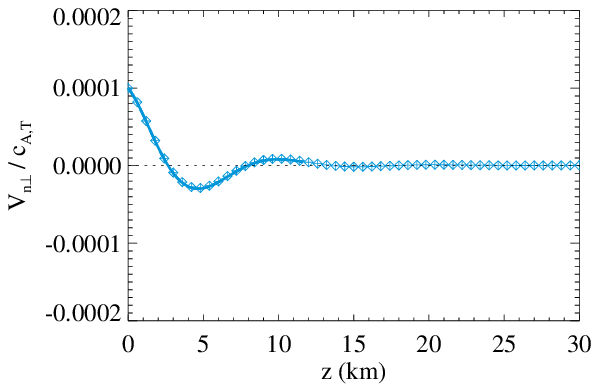}
        \includegraphics[width=0.32\hsize]{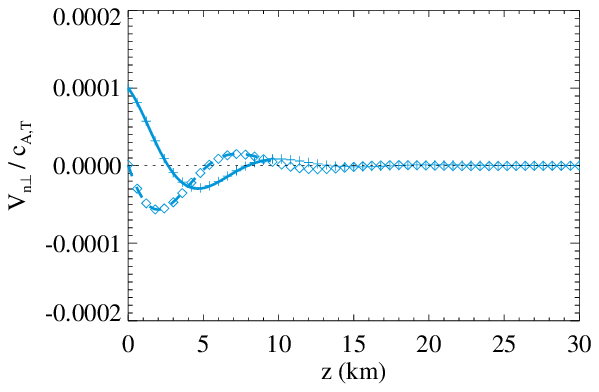}
        \includegraphics[width=0.32\hsize]{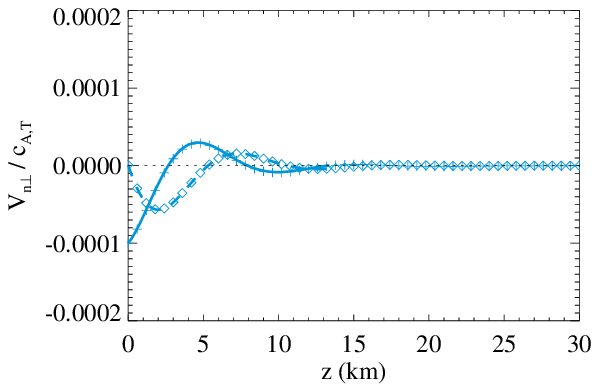} \\
        \includegraphics[width=0.32\hsize]{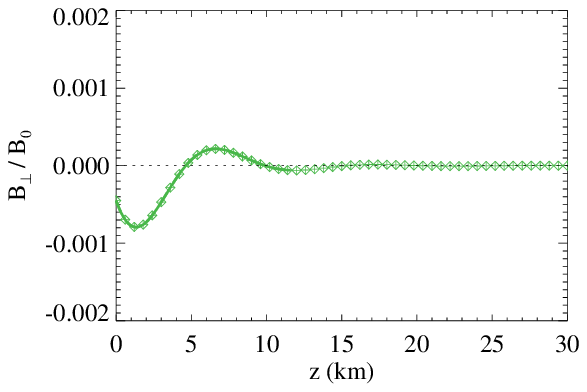}
        \includegraphics[width=0.32\hsize]{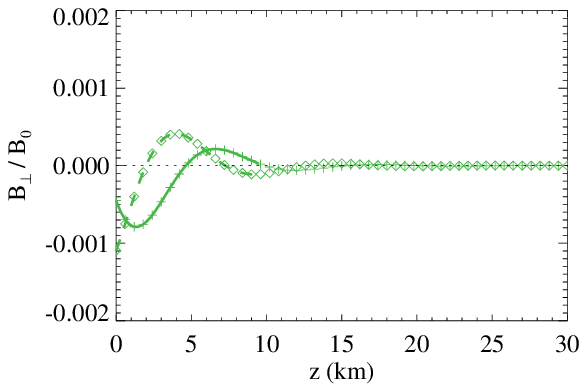}
        \includegraphics[width=0.32\hsize]{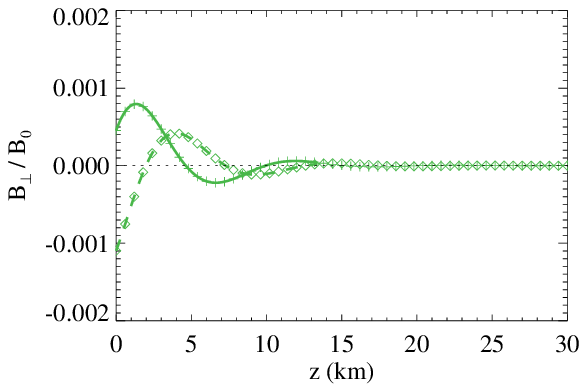}
        \caption{Same as Figure \ref{fig:firstorder_strong} but at the time $t = 10 \tau$ for a case with $\nu_{\rm{nc}} = 0.01 \omega$.}
        \label{fig:firstorder_weak}
    \end{figure*}
    
    The amplitudes and phase shifts $A_{\rm{n}}$, $A_{\rm{B}}$, $\phi_{\rm{n}}$, and $\phi_{\rm{B}}$ can be computed from Eqs. (\ref{eq:vnvc}) and (\ref{eq:b1vc}) as:
    \begin{equation} \label{eq:an_ab}
        A_{\rm{n}} = | \tilde{R}_{\rm{V}} | A_{\rm{c}}, \quad A_{\rm{B}} = | \tilde{R}_{\rm{B}} | A_{\rm{c}},
    \end{equation}

    \begin{equation} \label{eq:phin_phib}
        \phi_{\rm{n}} = \arg \left(\tilde{R}_{\rm{V}}\right) \quad \text{and} \quad \phi_{\rm{B}} = \arg \left( \tilde{R}_{\rm{B}} \right),
    \end{equation}
    after solving Eq. (\ref{eq:kz2_hall_gamma}) to obtain the value of $k_{z}$.

    To illustrate the different behavior of the linearly and circularly polarized modes, Figure \ref{fig:firstorder_strong} shows the results of applying the two-fluid model to a plasma with physical parameters typical of weakly ionized solar prominences. Here, we assume that the plasma is composed of hydrogen only, so we have that $m_{\rm{c}} = m_{\rm{n}} = m_{\rm{p}}$, where $m_{\rm{p}}$ is the proton mass, and $n_{\rm{e}} = n_{\rm{p}}$, where $n_{\rm{p}}$ is the number density of protons, so $n_{\rm{c}} = n_{\rm{e}} + n_{\rm{p}}= 2 n_{\rm{p}}$. For the present analysis we have chosen the following set of physical parameters: $n_{\rm{p}} = 1.4 \times 10^{15} \ \rm{m^{-3}}$, $n_{\rm{n}} = 1.4 \times 10^{17} \ \rm{m^{-3}}$, $B_{\rm{0}} = 10 \ \rm{G}$, $T_{\rm{c}}\suz = T_{\rm{n}}\suz = 10^{4} \ \rm{K}$. Therefore, we have that $\chi = 100$, $c_{\rm{A}} \approx 583 \ \rm{km \ s^{-1}}$, $\Omega_{i} \approx 95800 \ \rm{rad \ s^{-1}}$, and $\Omega_{\rm{H}} \approx 948 \ \rm{rad \ s^{-1}}$. Then, we consider the case of waves with a period of $\tau = 0.02$ (so $\omega \approx 0.33 \Omega_{\rm{H}}$) and a strong collisional coupling given by $\nu_{\rm{nc}} = 100 \omega$.

    The numerical simulations represented by the symbols in the left column of Figure \ref{fig:firstorder_strong} have been obtained by neglecting the contribution of Hall's term in the induction equation and applying at $z = 0$ the driving functions given by Eqs. (\ref{eq:1order_alfven}), in combination with the smoothing function given by Eq. (\ref{eq:smooth}). Thus, they correspond to the propagation of linearly polarized Alfvén waves. The amplitude of the driver for the velocity of the charged fluid is set as $A_{\rm{c}} = 0.01 c_{\rm{A,T}} \approx 0.58 \ \rm{km \ s^{-1}}$, where $c_{\rm{A,T}} = c_{\rm{A}}/\sqrt{1 + \chi}$ is the global or modified Alfvén speed, which takes into account the total density of the plasma. The other amplitudes ($A_{\rm{n}}$ and $A_{\rm{B}}$) and phase shifts ($\phi_{\rm{n}}$ and $\phi_{\rm{B}}$) are computed from Eqs. (\ref{eq:an_ab}) and (\ref{eq:phin_phib}) after solving the dispersion relation, Eq. (\ref{eq:kz2_hall_gamma}), for the chosen wave period and parameters of the plasma. The simulations run for a total time of 15 wave periods, so $t_{\rm{max}} = 0.30 \ \rm{s}$, and use a numerical domain of 2000 points to cover a distance of $20 \ \rm{km}$ (although here only half of the domain is shown to focus on the regions where the waves have reached the stationary stage and the influence from transient effects is negligible). Then, we see that the perturbations of the velocity of charges (top panel) and the velocity of neutrals (middle panel) propagate in phase while the perturbations of magnetic field (bottom panel) are in anti-phase with respect to the velocities, similarly to what occurs in the case of ideal MHD \citep{Walen1944ArMAF..30A...1W,Priest1984smh..book.....P}. In contrast to the ideal MHD scenario, the waves are affected by a weak damping caused by ion-neutral collisions, in agreement with previous analytical and numerical results on wave propagation in two-fluid partially ionized plasmas \citep[see, e.g.,][]{Kumar2003SoPh..214..241K,Mouschovias2011MNRAS.415.1751M,Zaqarashvili2011A&A...529A..82Z,Soler2013ApJ...767..171S,Kuzma2020A&A...639A..45K,Kraskiewicz2023SoPh..298...11K}. The solid lines included in these panels represent the analytical eigenfunctions given by Eqs. (\ref{eq:1order_alfven}) and show an excellent agreement with the numerical results.

    The centre and right columns display the results for the cases when the influence of Hall's current is taken into account. The simulations are performed by applying at $z = 0$ the driver given by Eqs. (\ref{eq:1order_circularly}), maintaining the same wave period and plasma parameters from the previous case. Here, we see that the $x$ and $y$ components of the perturbations do not overlap but there is a constant phase shift that depends on the polarization state. As for Alfvén waves, the velocities of the two fluids oscillate in phase while the magnetic field is in anti-phase. However, the damping rate and the wavenumber are larger for the ion-cyclotron mode and smaller for the whistler mode, showing how the effect of the collisional interaction strongly depends on the polarization state of the waves, as already discussed in \hyperlink{PaperI}{Paper I}. Another difference with respect to the case of Alfvén waves is that the normalized amplitude of the magnetic field perturbation is not the same as the normalized amplitude of the velocities. For the ion-cyclotron wave, the normalized amplitude of the magnetic field is smaller while for the whistler wave it is larger. This fact is related to the non-equipartition between the magnetic and kinetic energies of the waves \citep{Campos1992_10.1063/1.860136}, which is discussed in more detail in Section \ref{sec:wave_energy}.

    The waves represented in Figure \ref{fig:firstorder_strong} correspond to eigenmodes of the two-fluid system of equations (without and with Hall's term, respectively) and have been excited by applying very specific driving conditions. However, in a more general scenario the plasma would be affected by an arbitrary perturbation that does not correspond to an exact eigenmode, as in the configurations simulated by \citet{Kuzma2020A&A...639A..45K}, \citet{Kraskiewicz2023SoPh..298...11K}, and \citet{Kraskiewicz2025A&A...698A..74K}, for instance. Therefore, the resulting waves would be a combination of the previously described modes. To illustrate this behavior, we represent in Figure \ref{fig:firstorder_mix} the output from a simulation in which the driver for linear polarization, Eqs. (\ref{eq:1order_alfven}), has been applied but the effect of Hall's current has also been taken into account. This simulation uses 2000 points to cover a distance of 80 km and runs for a total time of 40 wave periods (so $t_{\rm{max}} = 0.80 \ \rm{s}$). Due to the different wavenumbers, phase speeds, and damping rates of the left-handed and right-handed modes, we see that the evolution of the velocity and magnetic field perturbations can be separated into two different regions. On the one hand, in $z \lesssim 30 \ \rm{km}$ we find evidences of the interaction of two oscillations modes with different wavenumbers. In this region, the perturbation would be described as the combination of a carrier wave and an envelope wave whose wavenumbers would be given by the half-sum and the half-difference of the wavenumbers of the ion-cyclotron and whistler modes. This behavior has already been shown, for instance, in simulations of propagating waves in resistive MHD \citep{Threlfall2012PhDT.......253T}, and in simulations of standing oscillations in multi-ion plasmas \citep{MartinezGomez2016ApJ...832..101M}. In addition, the damping of the perturbations is mostly determined by the damping rate of the ion-cyclotron mode (as shown by the black dashed curve). On the other hand, in the region $z > 30 \ \rm{km}$, we only find clear signs of one oscillation mode. The whistler mode has a larger phase speed than the ion-cyclotron mode and is less affected by ion-neutral collisions, so it can propagate to farther distances.

    Then, Figure \ref{fig:firstorder_weak} displays the results for a scenario with a weaker collisional coupling between the charged and neutral fluids than in the previous case. For this case, we set $\nu_{\rm{nc}} = 0.01 \omega$, the simulations run for a total time of $10$ wave periods, and the numerical domain covers a distance of $40 \ \rm{km}$ (although, once more, only a reduced section is shown in the plots). We see that there is a very pronounced damping due to the ion-neutral collisions. In addition, the velocities of each fluid are no longer in phase and the amplitude of the velocity of neutrals is much smaller than the amplitude of the velocity of charges. This is related to the low ionization degree of the plasma: due to the weak coupling, the motion of a small amount of charges is not able to generate an oscillation of the same magnitude in a much denser neutral fluid. In contrast with the results from Figure \ref{fig:firstorder_strong}, here the damping rate does not show a strong dependence on the polarization state of the waves and the three modes are almost completely attenuated in a distance of 2 to 3 wavelengths. The reason is that in a case of weak coupling, the difference in behavior caused by Hall's term becomes important for wave frequencies of the order of the ion cyclotron frequency, but here we have that $\omega \ll \Omega_{\rm{i}}$. Conversely, in Figure \ref{fig:firstorder_strong} larger differences were found between the ion-cyclotron and whistler modes because in a strong coupling scenario the reference frequency is the Hall frequency, $\Omega_{\rm{H}}$, which is much smaller than $\Omega_{\rm{i}}$.

    To conclude this section, we note that, in order to clearly illustrate how differently each polarization state is affected by the ion-neutral interaction, we have explored a particular set-up with waves of very high frequency, of the order of the Hall frequency ($\Omega_{\rm{H}}$). As a result we find damping lengths of the order of $10$ to $100 \ \rm{km}$. These values are similar to those obtained by \citet{Kuzma2020A&A...639A..45K} and \citet{Kraskiewicz2025A&A...698A..74K} for waves with periods shorter than $0.1 \ \rm{s}$ propagating at the intermediate layers of the solar chromosphere, although in those works a density ratio of $\chi \sim 1$ was considered and the influence of Hall's term was not taken into account. Therefore, these high-frequency (or short-period) waves would not be able to reach very far distances from the position of the driving source, and the plasma heating that they may produce would be confined to very small regions (as we show in more detail in Sections \ref{sec:wave_energy} and \ref{sec:nonlinear}). In contrast, it has already been shown that low-frequency waves have much larger damping lengths, which typically increase with the square of the wave period \citep[see, e.g.,][]{Zaqarashvili2011A&A...529A..82Z,Soler2013ApJ...767..171S}. In particular, the nonlinear two-fluid simulations performed by, e.g., \citet{Kuzma2020A&A...639A..45K}, \citet{Kraskiewicz2023SoPh..298...11K}, and \citet{Kraskiewicz2025A&A...698A..74K} have shown that low frequency Alfvén waves generated at the photosphere or the lower layers of the solar chromosphere are weakly affected by the collisional damping and can propagate to the upper layers where the damping becomes more efficient and they can produce a noticeable increase of the plasma temperature. Finally, the modulation of the wave amplitude due to the interaction of the ion-cyclotron and whistler modes illustrated in Figure \ref{fig:firstorder_mix} would not completely disappear in the low-frequency range but it would only become evident over very large distances.
    
\section{Results for second-order perturbations} \label{sec:second_order}
\subsection{Second-order equations} \label{sec:second_order_eqs}
    If we now gather all the terms proportional to $\epsilon^{2}$ in the perturbative expansion, we obtain the second-order equations for the perturbations in densities, pressures, and the longitudinal component of velocities. Defining $V_{\rm{n}\parallel} \equiv V_{\rm{n}z}\sut$ and $V_{\rm{c}\parallel} \equiv V_{\rm{c}z}\sut$, these equations are given by 
    \begin{equation} \label{eq:rho_s_2}
        \frac{\partial \rho_{\rm{s}}\sut}{\partial t} + \rho_{\rm{s0}} \frac{\partial V_{\rm{s}\parallel}}{\partial z} = 0, \quad s \in \{c, \ n\}
    \end{equation}

    \begin{equation} \label{eq:vnz_2}
        \rho_{\rm{n0}} \frac{\partial V_{\rm{n}\parallel}}{\partial t} =-\frac{\partial P_{\rm{n}}\sut}{\partial z} + \alpha_{\rm{cn}}\suz \left(V_{\rm{c}\parallel} - V_{\rm{n}\parallel} \right),
    \end{equation}
    
    \begin{equation} \label{eq:vcz_2}
        \rho_{\rm{c0}} \frac{\partial V_{\rm{c}\parallel}}{\partial t} =-\frac{\partial P_{\rm{c}}\sut}{\partial z} - \frac{\partial}{\partial z}\left(\frac{B_{\perp}^{2}}{2 \mu_{0}} \right) + \alpha_{\rm{cn}}\suz \left(V_{\rm{n}\parallel} - V_{\rm{c}\parallel} \right),
    \end{equation}

    \begin{equation} \label{eq:presn_2}
        \frac{\partial P_{\rm{n}}\sut}{\partial t} + \gamma P_{\rm{n0}} \frac{\partial V_{\rm{n}\parallel}}{\partial z} = \left(\gamma - 1 \right)Q_{\rm{nc}}\sut,
    \end{equation}

    \begin{equation} \label{eq:presc_2}
        \frac{\partial P_{\rm{c}}\sut}{\partial t} + \gamma P_{\rm{c0}} \frac{\partial V_{\rm{c}\parallel}}{\partial z} = \left(\gamma - 1 \right)Q_{\rm{cn}}\sut,
    \end{equation}
    where $P_{\rm{n0}}$ and $P_{\rm{c0}}$ are the background pressures of the neutral and charged fluid, respectively, and $\gamma$ is the adiabatic constant. The energy transfer terms due to elastic collisions are given by \citep[see, e.g.,][]{Schunk1977RvGSP..15..429S,Draine1986MNRAS.220..133D}
    \begin{equation} \label{eq:qnc2}
        Q_{\rm{nc}}\sut = \frac{\alpha_{\rm{cn}}\suz}{m_{\rm{n}} + m_{\rm{c}}} \left[3k_{\rm{B}} \Delta T\sut +  m_{\rm{c}}V_{D}^2\right]
    \end{equation}
    and
    \begin{equation} \label{eq:qcn2}
        Q_{\rm{cn}}\sut = \frac{\alpha_{\rm{cn}}\suz}{m_{\rm{c}} + m_{\rm{n}}} \left[-3k_{\rm{B}} \Delta T\sut + m_{\rm{n}}V_{\rm{D}}^2\right],
    \end{equation}
    with 
    \begin{equation} \label{eq:deltat2_vd}
        \Delta T\sut \equiv T_{\rm{c}}\sut - T_{\rm{n}}\sut \quad \text{and} \quad V_{\rm{D}}^{2} \equiv \left(\bm{V}_{\rm{c}\perp} - \bm{V}_{\rm{n}\perp} \right)^{2},
    \end{equation}
    where $T_{\rm{c}}\sut$ and $T_{\rm{n}}\sut$ are the second-order perturbations of temperature, and $k_{\rm{B}}$ is the Boltzmann constant.

    The system of equations (\ref{eq:rho_s_2}) - (\ref{eq:presc_2}) shows that the second-order longitudinal perturbations are not coupled to the second-order transverse perturbations. However, they depend on the first-order transverse perturbations through the gradient of the magnetic pressure, as it already occurs in fully ionized plasmas \citep[see, e.g.,][]{Hollweg1971JGR....76.5155H,Cramer2001paw..book.....C}, and through the energy transfer terms $Q_{\rm{nc}}\sut$ and $Q_{\rm{cn}}\sut$.

\subsection{Wave energy and heating rate} \label{sec:wave_energy}
    The internal energy of a fluid is related to its pressure by $e_{\rm{s}} = P_{\rm{s}}/\left(\gamma - 1 \right)$. Hence, from Eqs. (\ref{eq:presn_2}) and (\ref{eq:presc_2}), we can get that the total internal energy of the plasma increases at the heating rate
    \begin{equation} \label{eq:heating}
        \mathcal{H} = Q_{\rm{cn}}\sut + Q_{\rm{nc}}\sut = \alpha_{\rm{cn}}\suz V_{\rm{D}}^{2}.
    \end{equation}

    Following \citet{Watanabe1961CaJPh..39.1044W}, \citet{Braginskii1965RvPP....1..205B}, \citet{Walker2004SPP....16.....W} or \citet{Soler2016A&A...592A..28S}, it is possible to obtain a second-order evolution equation for the wave energy associated with the first-order perturbations. This wave energy equation has the general form
    \begin{equation} \label{eq:wave_energy}
        \frac{\partial \mathcal{U}}{\partial t} + \nabla \cdot \bm{\Pi} = - \mathcal{L},
    \end{equation}
    where $\mathcal{U}$ is the total energy density of the waves, $\bm{\Pi}$ is the energy flux, and $\mathcal{L}$ is the energy loss rate.

    Using Eqs. (\ref{eq:vn1_perp}) - (\ref{eq:by1}) to compute the corresponding wave energy equation, we get that the total wave energy density, $\mathcal{U}$, is the sum of the kinetic energy of charges, $\mathcal{K}_{\rm{c}}$, the kinetic energy of neutrals, $\mathcal{K}_{\rm{n}}$, and the magnetic energy, $\mathcal{M}$, with
    \begin{equation} \label{eq:energies}
        \mathcal{K}_{\rm{s}} = \frac{\rho_{\rm{s0}}\bm{V}_{\rm{s}\perp}^{2}}{2} \ \text{for} \ s \in \{c,n\} \ \text{and} \ \mathcal{M} = \frac{\bm{B}_{\perp}^{2}}{2 \mu_{\rm{0}}}.
    \end{equation}
    We also find that $\mathcal{L} = \mathcal{H}$, with $\mathcal{H}$ given by Eq. (\ref{eq:heating}), showing that the energy lost by the waves due to the collisional interaction leads to an increase of the internal energy of the plasma \citep[see, e.g.,][]{Braginskii1965RvPP....1..205B,Draine1986MNRAS.220..133D}.

    If Hall's term is included in the induction equation, the complete expression for the wave energy flux is $\bm{\Pi} = \bm{\Pi}_{\rm{adv}} + \bm{\Pi}_{\rm{Hall}}$, with
    \begin{eqnarray} \label{eq:flux_adv}
        \bm{\Pi}_{\rm{adv}} = \frac{1}{\mu_{\rm{0}}} \left[\left(\bm{B}_{\rm{0}} \cdot \bm{B}_{\perp} \right) \bm{V}_{\rm{c}\perp} - \left(\bm{V}_{\rm{c}\perp} \cdot \bm{B}_{\perp} \right) \bm{B}_{\rm{0}} \right]  
    \end{eqnarray}
    and
    \begin{equation}
        \bm{\Pi}_{\rm{Hall}} = \frac{1}{e n_{\rm{e0}} \mu_{\rm{0}}} \left[ \left(\bm{B}_{\perp} \cdot \bm{J}\suo \right) \bm{B}_{\rm{0}} - \left(\bm{B}_{\perp} \cdot \bm{B}_{\rm{0}} \right) \bm{J}\suo \right],
    \end{equation}
    where $\bm{J}\suo = \left(\nabla \times \bm{B}\suo \right) / \mu_{\rm{0}}$ is the first-order perturbation of the current density. We note here that in \hyperlink{PaperI}{Paper I} it was incorrectly stated that the contribution of $\bm{\Pi}_{\rm{Hall}}$ generally vanishes for first-order Alfvénic waves. That statement is indeed correct for linearly polarized waves, which fulfill that $\bm{B}_{\perp} \cdot \bm{J}\suo = 0$ and $\bm{B}_{\perp} \cdot \bm{B}_{\rm{0}} = 0$, but it does not apply for circularly polarized waves, which have $\bm{B}_{\perp} \cdot \bm{J}\suo \neq 0$. Therefore, the contribution of Hall's term to the energy flux should not be straightforwardly neglected.

    Now, we focus on the expression for the heating rate $\mathcal{H}$. From Eq. (\ref{eq:vnvc}) we can obtain that the time-averaged perturbations of velocity of neutrals and charges are related by
    \begin{equation} \label{eq:vnvc2}
        \langle\bm{V}_{\rm{n}\perp}^{2} \rangle = \frac{\nu_{\rm{nc}}^{2}}{\omega^{2} + \nu_{\rm{nc}}^{2}} \langle\bm{V}_{\rm{c}\perp}^{2}\rangle.
    \end{equation}
    Then, combining Eqs. (\ref{eq:heating}) and (\ref{eq:vnvc2}), we can write the time-averaged heating rate as
    \begin{equation} \label{eq:hrate_av}
        \langle \mathcal{H}\rangle = \rho_{\rm{c0}} \nu_{\rm{cn}} \frac{\omega^{2}}{\nu_{\rm{nc}}^{2} + \omega^{2}}\langle\bm{V}_{\rm{c}\perp}^{2} \rangle.
    \end{equation}
    This expression is equivalent to Equation (8) of \citet{Song2011JGRA..116.9104S} (if the contribution from resistivity is not taken into account in that equation). However, for the purposes of this work it is more useful to note that $\rho_{\rm{c0}} \langle\bm{V}_{\rm{c}\perp}^{2}\rangle = 2 \langle \mathcal{K}_{\rm{c}} \rangle$, and to apply the expressions derived in \hyperlink{PaperI}{Paper I} for the relations between the total kinetic energy ($\mathcal{K}_{\rm{T}} = \mathcal{K}_{\rm{c}} + \mathcal{K}_{\rm{n}}$), the magnetic energy and the kinetic energy of charges, namely 
    \begin{equation} \label{eq:ek1_tot}
        \langle \mathcal{K}_{\rm{T}} \rangle = \langle \mathcal{K}_{\rm{c}} \rangle \left[1 + \frac{\chi \nu_{\rm{nc}}^{2}}{\omega^{2} + \nu_{\rm{nc}}^{2}} \right] = \langle \mathcal{K}_{\rm{c}} \rangle \left[\frac{\omega^{2} + \left(1 + \chi \right) \nu_{\rm{nc}}^{2}}{\omega^{2} + \nu_{\rm{nc}}^{2}} \right]
    \end{equation}
    and
    \begin{equation} \label{eq:eb1_hall}
        \langle \mathcal{M} \rangle = \langle \mathcal{K}_{\rm{c}} \rangle \sqrt{\frac{\left(\omega^{2} \Gamma_{\rm{i}}^{2} + \nu_{\rm{nc}}^{2} \Gamma_{\rm{H}}^{2} \right) \left(\omega^{2} + \left(1 + \chi \right)^{2} \nu_{\rm{nc}}^{2} \right)}{\left(\omega^{2} + \nu_{\rm{nc}}^{2} \right)^{2}}},
   \end{equation}
    where Eq. (\ref{eq:eb1_hall}) is a simplified (but equivalent) version of Eq. (65) from \hyperlink{PaperI}{Paper I}, which can be obtained from the combination of Eqs. (\ref{eq:kz2_hall_gamma}), (\ref{eq:b1vc}) and (\ref{eq:energies}).  

    \begin{figure*}
        \includegraphics[width=0.32\hsize]{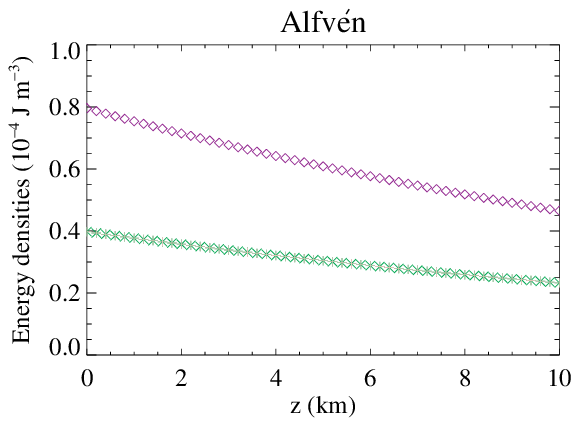}
        \includegraphics[width=0.32\hsize]{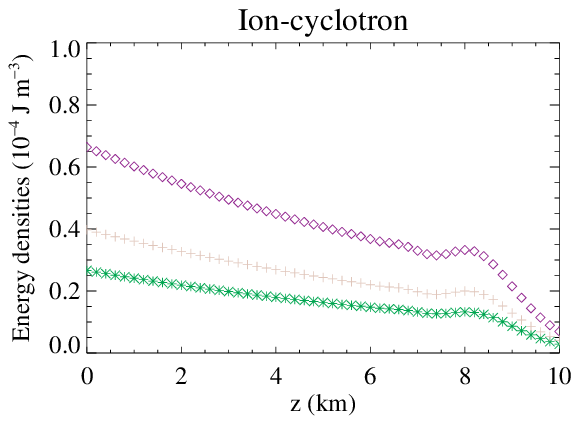}
        \includegraphics[width=0.32\hsize]{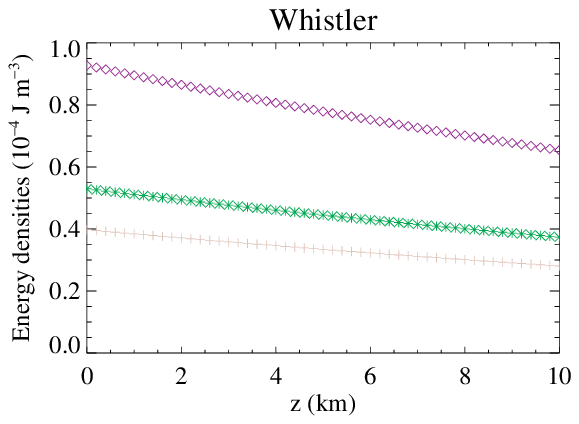} \\
        \includegraphics[width=0.32\hsize]{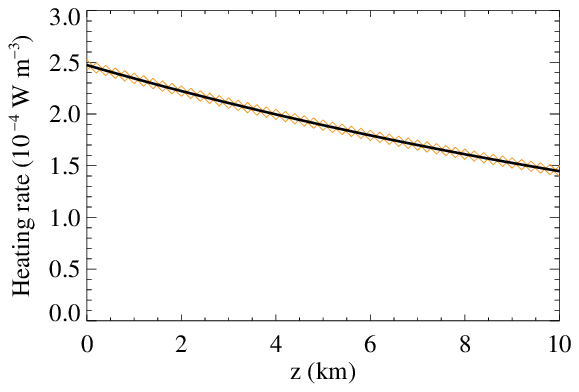}
        \includegraphics[width=0.32\hsize]{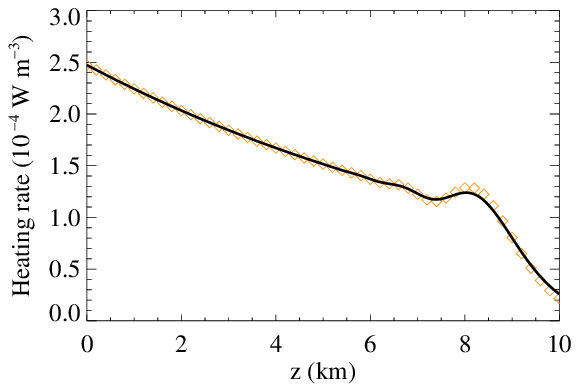}
        \includegraphics[width=0.32\hsize]{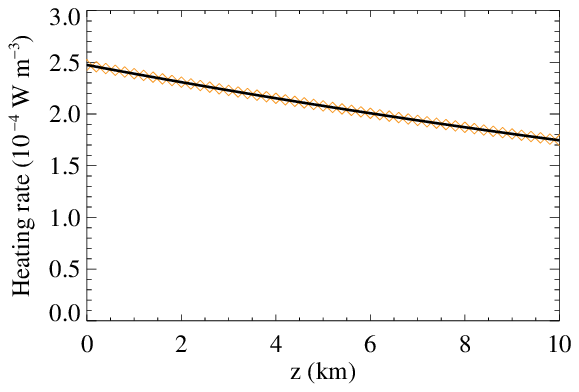} \\
        \caption{Time-averaged energy densities (top panels) and heating rates (bottom panels) computed from simulations of Alfvén waves (left column), ion-cyclotron waves (centre column) and whistler waves (right column) with a period of $\tau = 0.02$ and amplitude of the driver given by $A_{\rm{c}} = 0.01 c_{\rm{A,T}}$. Case of a plasma with $n_{\rm{p}} = 1.4 \times 10^{15} \ \rm{m^{-3}}$, $\chi = 100$, $B_{\rm{0}} = 10 \ \rm{G}$, and $\nu_{\rm{nc}} = 100 \omega$. Diamonds, asterisks, and crosses in the top panels represent the total energy density $\langle \mathcal{U} \rangle$, the magnetic energy density $\langle \mathcal{M} \rangle$, and the total kinetic energy density $\langle \mathcal{K} \rangle$. The solid lines in the bottom panels represent the heating rates computed using Eq. (\ref{eq:h_to_u}).}
        \label{fig:energies_strong}
    \end{figure*}
    
    In this way, we rewrite the time-averaged heating rate in terms of the total energy density, $\mathcal{U} = \mathcal{K}_{\rm{T}} + \mathcal{M}$, as
    \begin{equation}
        \langle \mathcal{H} \rangle = \frac{2 \chi \nu_{\rm{nc}} \omega^{2}}{\omega^{2} + \left(1 + \chi \right) \nu_{\rm{nc}}^2 + \Theta} \langle \mathcal{U} \rangle,
        \label{eq:h_to_u}
    \end{equation}
    with
    \begin{equation} \label{eq:aux_h}
        \Theta \equiv \left[\left(\omega^{2} \Gamma_{\rm{i}}^{2} + \nu_{\rm{nc}}^{2} \Gamma_{\rm{H}}^{2} \right) \left(\omega^{2} + \left(1 + \chi \right)^{2} \nu_{\rm{nc}}^{2} \right) \right]^{1/2}.
    \end{equation}

    If we compare Eq. (\ref{eq:h_to_u}) with the expression for the quality factor given by Eq. (\ref{eq:qfactor_exact}), we can notice that in the limit of low frequencies, when the effect of Hall's term can be neglected and $\{\Gamma_{\rm{i}}, \Gamma_{\rm{H}} \} \approx 1$, the following relation is fulfilled:
    \begin{equation} \label{eq:hqu}
        \langle \mathcal{H} \rangle = \frac{\omega}{Q\left(\omega \right)} \langle \mathcal{U} \rangle,
    \end{equation}
    where we have written the quality factor as $Q\left(\omega \right)$ to specifically remark that this parameter is not a constant but a function of the wave frequency. The above expression relating the rate of energy dissipation, the total energy density and the quality factor commonly appears in the field of engineering \citep[see, e.g.,][]{Gustaffsson2005,Christopoulos2019OExpr..2714505C}. In addition, equivalent relations have been previously obtained in the field of plasma physics for the case of magnetohydrodynamic (MHD) waves generated by impulsive drivers, where the wavenumber has been assumed real and $\omega = \omega_{\rm{R}} + i \omega_{\rm{I}}$, with $\omega_{\rm{R}}$ the oscillation frequency and $\omega_{\rm{I}}$ the damping rate, and the quality factor is then given by $Q = |\omega_{\rm{R}}| / (2|\omega_{\rm{I}}|)$. For instance, \citet{Braginskii1965RvPP....1..205B}, \citet{Tracy2014rtb..book.....T} or \citet{CallyGomezMiguez2023ApJ...946..108C} have shown that if the damping is weak ($|\omega_{\rm{I}}| \ll |\omega_{\rm{R}}|$), the heating rate can be computed as $\mathcal{H} = 2 |\omega_{\rm{I}}| \mathcal{U} \approx (\omega / Q) \mathcal{U} $. Nonetheless, we note that here no restriction has been imposed on the strength of the damping for the derivation of Eq. (\ref{eq:hqu}), so it has a larger range of applicability and it remains valid in the regime of strong damping. However, in the range of high frequencies, where $\omega \sim \Omega_{\rm{i}}$ or $\omega \sim \Omega_{\rm{H}}$, Eq. (\ref{eq:hqu}) loses its accuracy. This is consistent with previous findings about propagation of waves in dispersive media \citep{Gustafsson2017RaSc...52.1325G,Christopoulos2019OExpr..2714505C}.
    
    \begin{figure*}
        \includegraphics[width=0.32\hsize]{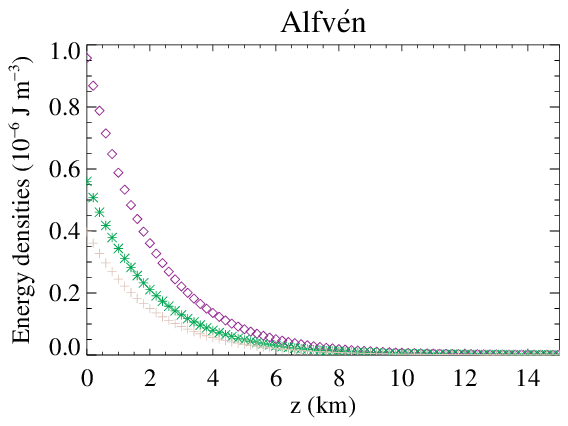}
        \includegraphics[width=0.32\hsize]{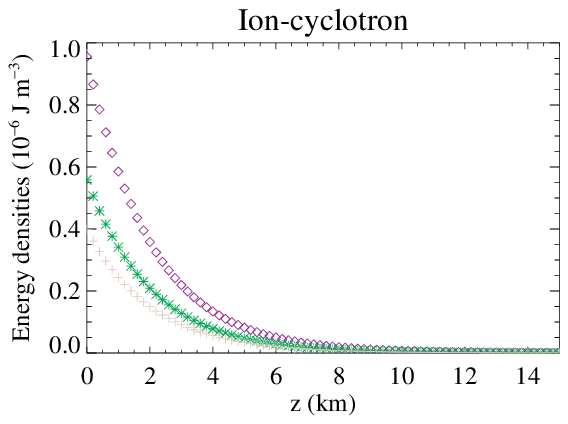}
        \includegraphics[width=0.32\hsize]{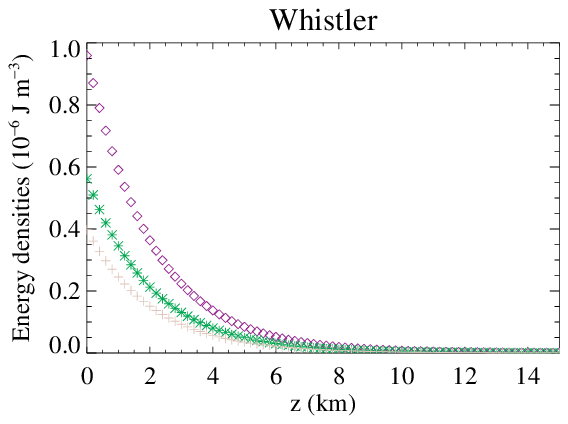} \\
        \includegraphics[width=0.32\hsize]{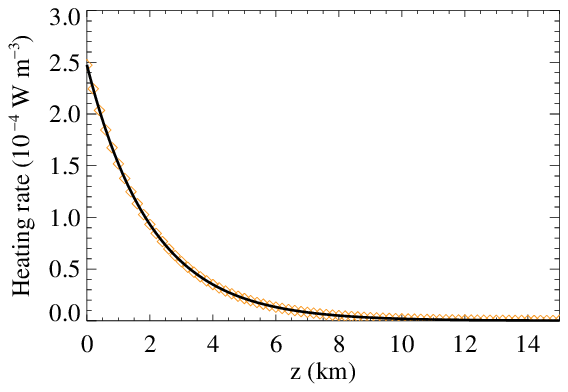}
        \includegraphics[width=0.32\hsize]{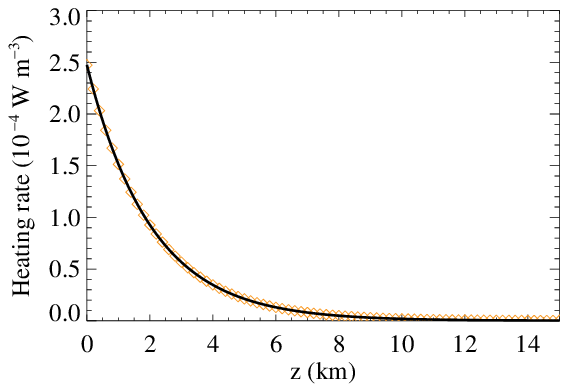}
        \includegraphics[width=0.32\hsize]{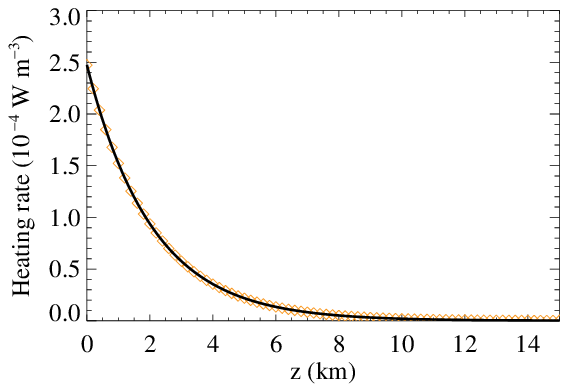} \\
        \caption{Same as Figure \ref{fig:energies_strong} but for a case with $\nu_{\rm{nc}} = 0.01 \omega$.}
        \label{fig:energies_weak}
    \end{figure*}

    To check the validity of the previous analytic computations, we show in Figures \ref{fig:energies_strong} and \ref{fig:energies_weak} time-averages of the kinetic energy, magnetic energy and total energy densities (top panels) and heating rates (bottom panels) from the simulations represented in Figures \ref{fig:firstorder_strong} and \ref{fig:firstorder_weak}. In order to ensure that the results correspond to the stationary stage of the waves and to improve their accuracy, the time-averages are computed over the last three periods of the simulations.
    
    The top left panel of Figure \ref{fig:energies_strong} evidences that there is equipartition between the kinetic energy and the magnetic energy of Alfvén waves in the case with a strong collisional coupling. However, the centre and right panels show that the ion-cyclotron modes have more kinetic energy than magnetic energy, while the opposite occurs for whistler waves, in agreement with the analysis performed by \citet{Campos1992_10.1063/1.860136} and in \hyperlink{PaperI}{Paper I}. In the simulations considered here, the driver provides the same kinetic energy at $z = 0$ for the three modes, but the total energy varies with the polarization state. This explains why in the bottom panels of Figure \ref{fig:energies_strong} the heating rates  $\langle \mathcal{H} \rangle$ of the whistler mode are larger in most of the numerical domain than those of the ion-cyclotron mode although the latter is affected by stronger damping rates ($k_{\rm{I}}$). At $z = 0$, the three modes have the same heating rate because it directly depends on the kinetic energy, as can be deduced from Eq. (\ref{eq:hrate_av}), but this amount of energy dissipated by collisions represents a larger fraction of the total wave energy in the case of the ion-cyclotron mode than in the whistler mode. Therefore, as the waves propagate, the ion-cyclotron mode has less energy available to be dissipated. If we have otherwise considered a case in which the three modes started with the same total energy, the average heating rates would be initially larger for the ion-cyclotron mode, since it has a smaller quality factor. In addition, the bottom panels of Figure \ref{fig:energies_strong} show an excellent agreement between the heating rates computed from the simulations using Eq. (\ref{eq:hrate_av}) directly (orange symbols) and those computed after applying Eq. (\ref{eq:h_to_u}) to the total wave energy density (black solid lines).

    Then, Figure \ref{fig:energies_weak} shows the results from the simulations with a weak collisional coupling. The wave energy is almost completely dissipated in a very short distance and there are no remarkable differences between the three polarization states. Moreover, the magnetic energy dominates over the kinetic energy of the waves, in agreement with the analytical computations from \hyperlink{PaperI}{Paper I}.

\subsection{Nonlinearly generated perturbations} \label{sec:nonlinear}
\begin{figure*}
        \includegraphics[width=0.32\hsize]{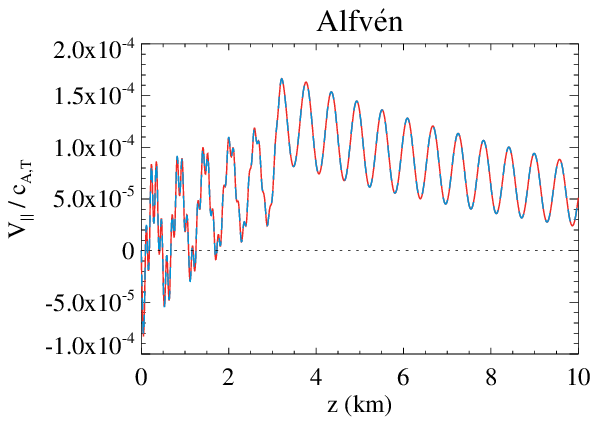}
        \includegraphics[width=0.32\hsize]{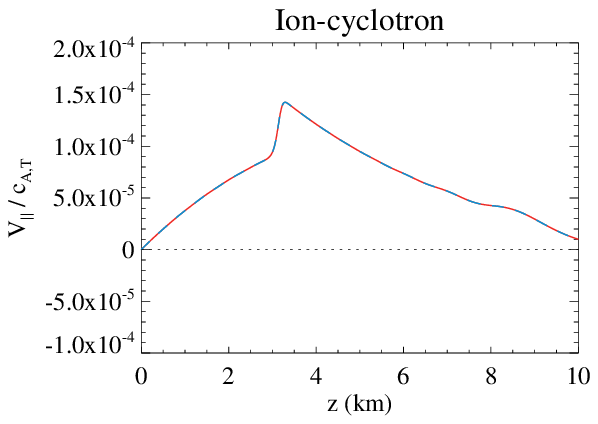}
        \includegraphics[width=0.32\hsize]{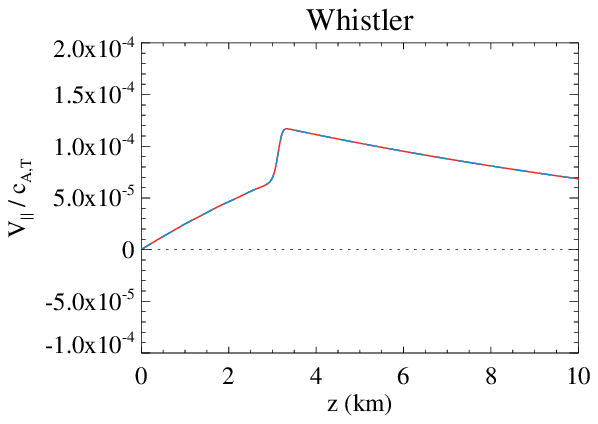} \\
        \includegraphics[width=0.32\hsize]{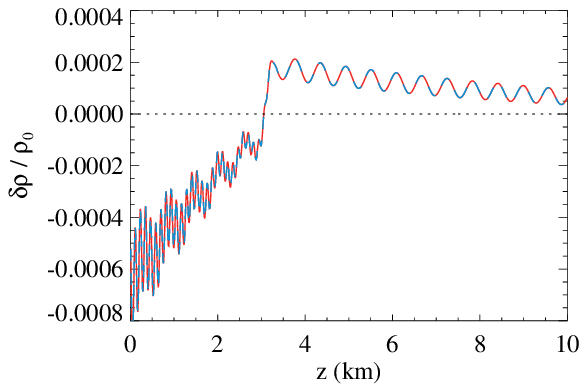}
        \includegraphics[width=0.32\hsize]{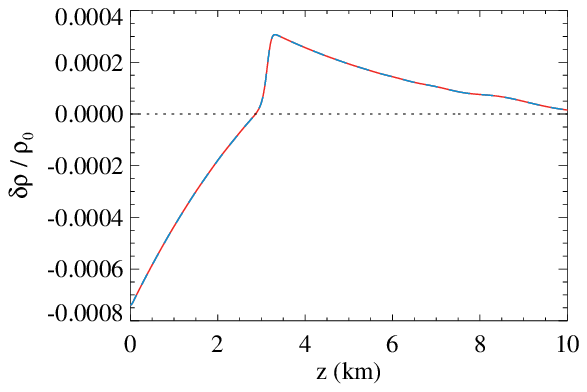}
        \includegraphics[width=0.32\hsize]{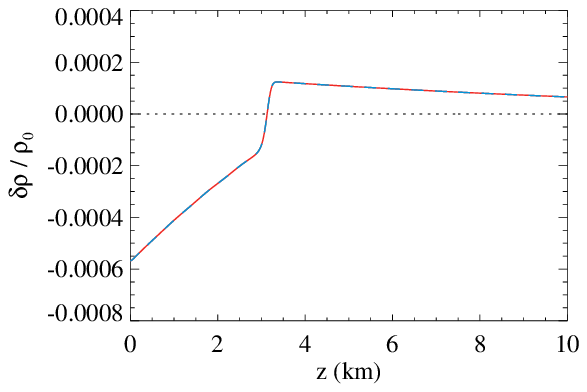} \\
        \includegraphics[width=0.32\hsize]{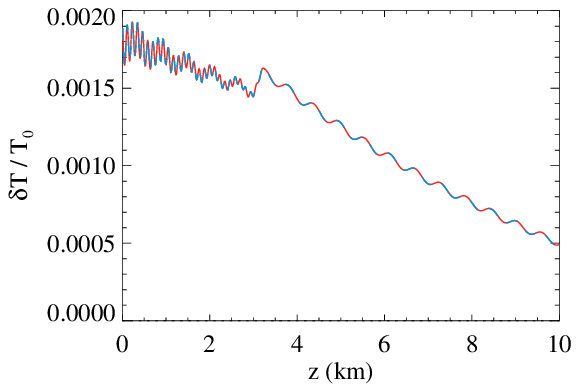}
        \includegraphics[width=0.32\hsize]{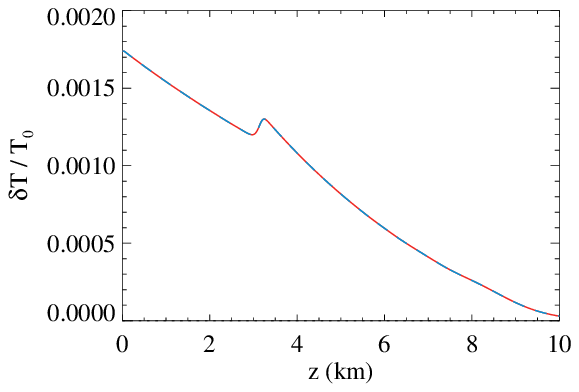}
        \includegraphics[width=0.32\hsize]{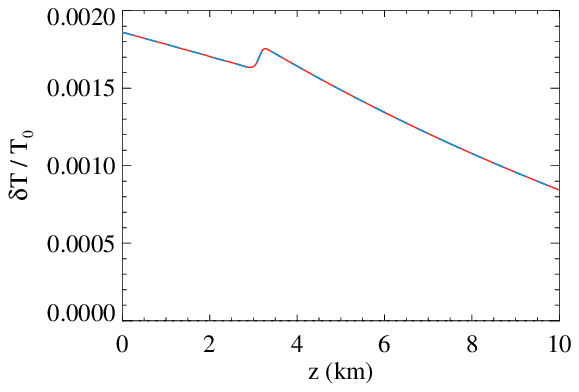}
        \caption{Normalized perturbations of the longitudinal component of velocity (top row), density (middle row), and temperature (bottom row) from simulations of Alfvén waves (left column), ion-cyclotron waves (centre column), and whistler waves (right column) with a period of $\tau = 0.02$ and driving amplitude $A_{\rm{c}} = 0.01 \ \rm{c_{\rm{A,T}}}$. Case of a plasma with $n_{\rm{p}}=1.4 \times 10^{15} \ \rm{m^{-3}}$, $\chi = 100$, $B_{\rm{0}} = 10 \ \rm{G}$, and $\nu_{\rm{nc}} = 100 \omega$. The red solid lines correspond to the results for the charged fluid, while the blue dashed lines represent the results for the neutral fluid, computed at the time $t = 15 \tau$.}
        \label{fig:secondorder_strong}
    \end{figure*}

    In this section we analyze the properties of the second-order perturbations generated by the transverse waves discussed in Section \ref{sec:eigenfunction}. We start by describing the results of the numerical simulations for the case of strong collisional coupling. The panels in the top row of Figure \ref{fig:secondorder_strong} display the normalized longitudinal component of the velocity; the panels in the middle row show the normalized perturbations of density, $\delta \rho_{\rm{s}} / \rho_{\rm{s0}} \equiv \left(\rho_{\rm{s}}\sut - \rho_{\rm{s0}} \right) / \rho_{\rm{s0}}$; and the panels at the bottom row display the normalized perturbations of temperature, $\delta T_{\rm{s}} / T_{\rm{s0}} \equiv \left(T_{\rm{s}}\sut - T_{\rm{s0}}\right) / T_{\rm{s0}}$.

    We first focus on the left panels of Figure \ref{fig:secondorder_strong}, which represent the results for linearly polarized Alfvén waves (without the influence of Hall's current). All the three panels show signatures of the propagation of various waves with different phase speeds and wavenumbers. It is known from previous studies \citep[see, e.g.,][]{Hollweg1971JGR....76.5155H,Rankin1994JGR....9921291R,Zheng2016PhyS...91a5601Z} that in fully ionized plasmas the second-order perturbations generated by Alfvén waves are a combination of two waves with double the frequency of the original driver and amplitudes that depend quadratically on the amplitude of the first-order perturbations: one of the waves propagates at the Alfvén speed while the other one propagates at the sound speed. In partially ionized plasmas with strong collisional coupling the waves propagate at the global Alfvén speed, $c_{\rm{A,T}}$, and at an effective sound speed, $c_{\rm{eff}}$, respectively \citep{MartinezGomez2018ApJ...856...16M,Ballester2024RSPTA.38230222B}. For the two-fluid plasma considered in this work, the effective sound speed is given by
    \begin{equation} \label{eq:ceff}
        c_{\rm{eff}} = \left(\frac{c_{\rm{c}}^2 + \chi c_{\rm{n}}^{2}}{1 + \chi} \right)^{1/2},
    \end{equation}
    where $c_{\rm{s}} = \sqrt{\gamma P_{\rm{s0}}/\rho_{s0}}$ is the sound speed of the fluid ``s''. Due to the chosen plasma parameters, we have that $c_{\rm{eff}} \approx 11.8 \ \rm{km \ s^{-1}}$. Since $c_{\rm{eff}} < c_{\rm{A,T}}$, the region $0 < z \leq c_{\rm{eff}}t$ is affected by both nonlinearly generated waves, while the region $c_{\rm{eff}} t \leq z \leq c_{\rm{A,T}} t$ is only affected by the wave propagating at the global Alfvén speed.
    
    The presence of dissipative mechanisms, such as resistivity and viscosity \citep{McLaughlin2011A&A...527A.149M,Threlfall2012PhDT.......253T,Zheng2016PhyS...91c5601Z} or ion-neutral collisions \citep{MartinezGomez2018ApJ...856...16M,Ballester2024RSPTA.38230222B}, produces a bulk flow that displaces the plasma from the position where the original driver is applied. As seen in the top left panel of Figure \ref{fig:secondorder_strong}, there is a mostly positive longitudinal component of the velocity. In addition, the middle left panel shows a decrease of the plasma density in the region $z \leq c_{\rm{eff}}t$ but an increase of density in the region $z \geq c_{\rm{eff}}t$. The bulk flow is caused by the pressure gradients that are generated as the energy of the first-order propagating wave is dissipated and the temperature of the plasma is increased, as it can be seen in the bottom left panel.

    Then, we examine the centre and right columns of Figure \ref{fig:secondorder_strong}, which represent the results from the simulations that include Hall's term. There are no traces of the oscillatory motion that appears in the case of Alfvén waves, but the bulk flow is still present. Moreover, there are slight differences between the second-order perturbations associated to the ion-cyclotron mode and those associated to the whistler mode: the former shows a larger maximum amplitude of the bulk flow, which is related to the stronger gradients caused by a larger damping rate, while the latter produces a larger increment of the plasma temperature, which is explained by the difference in the total energy carried by the waves (see Figure \ref{fig:energies_strong}).

    To understand the origin of the main differences between the second-order perturbations associated to the linearly and to the circularly polarized modes, we come back to Eqs. (\ref{eq:rho_s_2}) - (\ref{eq:presc_2}). Taking into account the results for the temperature perturbations represented in the bottom panels of Figure \ref{fig:secondorder_strong}, we can neglect the terms related to the difference of temperatures, $\Delta T\sut$, in the expression for the energy transfer due to collisions, Eqs. (\ref{eq:qnc2}) and (\ref{eq:qcn2}). Thus, the evolution equations of pressures and the longitudinal velocities are not coupled to the equations for the density perturbations. If we combine Eqs. (\ref{eq:vnz_2}) - (\ref{eq:presc_2}) and apply the definitions of heating rate and magnetic energy density given by Eqs. (\ref{eq:heating}) and (\ref{eq:energies}), we obtain the following fourth-order differential equation for the longitudinal velocity of the charged fluid:
    \begin{gather}
        \mathcal{D}_{\parallel} V_{\rm{c\parallel}} = -\frac{1}{\rho_{\rm{c0}}} \Big[\partial_{t}^{2} + \nu_{\rm{nc}} \partial_{t} - c_{\rm{n}}^{2}\partial_{z}^{2} \Big]\partial_{t}\partial_{z} \mathcal{M} \nonumber \\
        -\frac{\left(\gamma - 1 \right)}{\rho_{\rm{c0}}}\Big\{\frac{m_{\rm{n}}}{m_{\rm{n}} + m_{\rm{c}}} \Big[\partial_{t}^{2} - c_{\rm{n}}^{2} \partial_{z}^{2} \Big] + \nu_{\rm{nc}} \partial_{t}\Big\}\partial_{z}\mathcal{H},
        \label{eq:vcz_ponder}
    \end{gather}
    where we have defined $\partial_{t} \equiv \frac{\partial}{\partial t}$ and $\partial_{z} \equiv \frac{\partial}{\partial z}$, and 
    \begin{gather}
        \mathcal{D}_{\parallel} \equiv \partial_{t}^{4} + \left(1 + \chi \right) \nu_{\rm{nc}} \partial_{t}^{3} - \left(c_{\rm{n}}^{2} + c_{\rm{c}}^{2} \right) \partial_{t}^{2}\partial_{z}^{2} \nonumber \\
        -\nu_{\rm{nc}} \left(c_{\rm{c}}^{2} + \chi c_{\rm{n}}^{2} \right) \partial_{t}\partial_{z}^{2} + c_{\rm{n}}^{2} c_{\rm{c}}^{2} \partial_{z}^{4}.
    \end{gather}

    A similar equation can be obtained for the neutral fluid:
    \begin{gather}
        \mathcal{D}_{\parallel} V_{\rm{n\parallel}} = - \frac{\nu_{\rm{nc}}}{\rho_{\rm{c0}}}\partial_{t}^{2} \partial_{z} \mathcal{M} -\frac{\left(\gamma - 1 \right)\nu_{\rm{nc}}}{\rho_{\rm{c0}}}\partial_{t}\partial_{z}\mathcal{H} \nonumber \\
        - \frac{\left(\gamma - 1 \right) m_{\rm{c}}}{\rho_{\rm{n}}\suz \left(m_{\rm{n}} + m_{\rm{c}} \right)}\Big[\partial_{t}^{2}-c_{\rm{c}}^{2} \partial_{z}^{2} \Big]\partial_{z}\mathcal{H}. 
        \label{eq:vnz_ponder}
    \end{gather}

    Then, Eq. (\ref{eq:rho_s_2}) can be used to compute the second-order perturbations of density after obtaining the longitudinal velocities from Eqs. (\ref{eq:vcz_ponder}) and (\ref{eq:vnz_ponder}).

     \begin{figure*}
        \includegraphics[width=0.32\hsize]{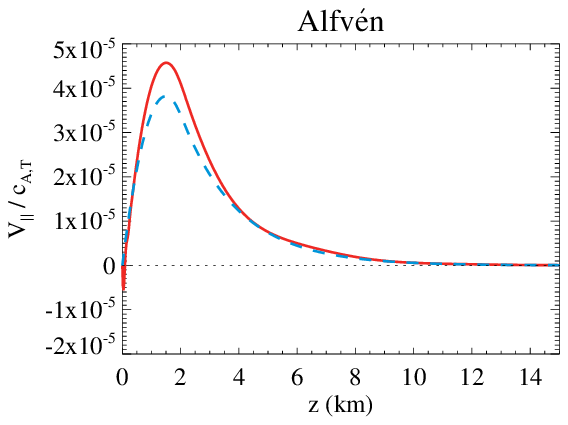}
        \includegraphics[width=0.32\hsize]{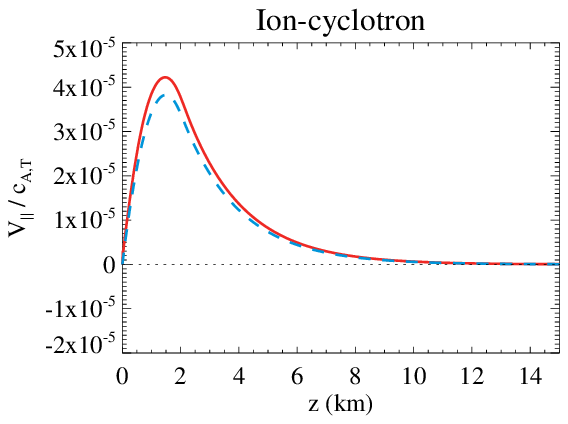}
        \includegraphics[width=0.32\hsize]{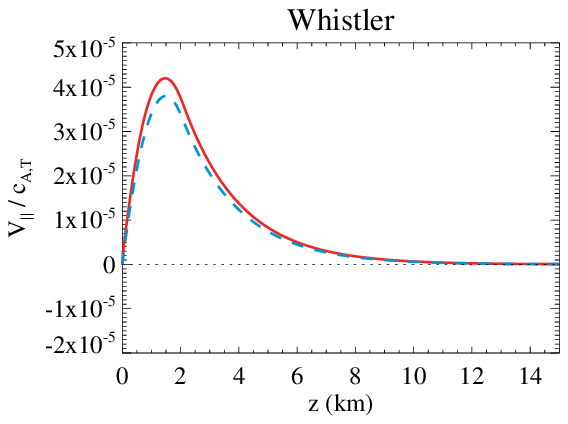} \\
        \includegraphics[width=0.32\hsize]{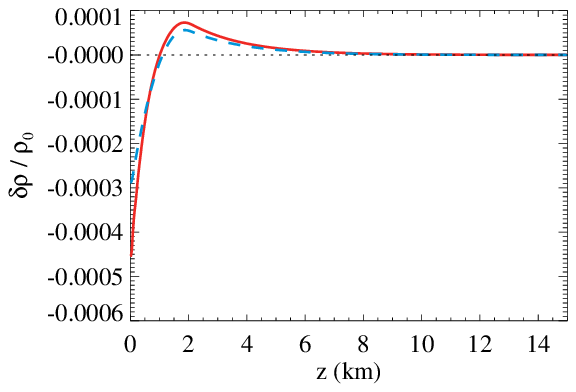}
        \includegraphics[width=0.32\hsize]{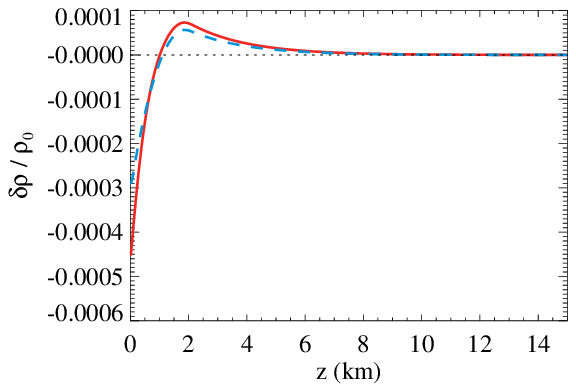}
        \includegraphics[width=0.32\hsize]{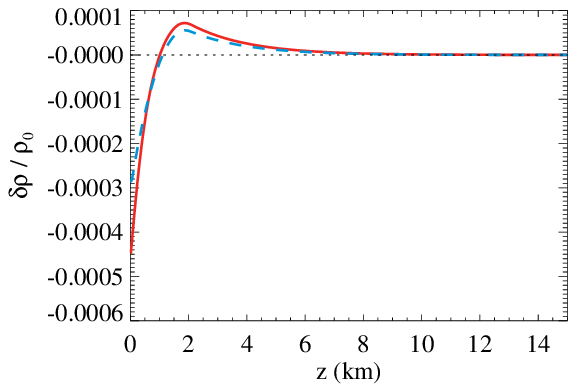} \\
        \includegraphics[width=0.32\hsize]{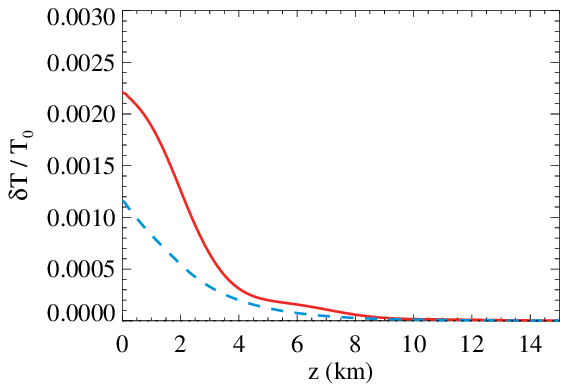}
        \includegraphics[width=0.32\hsize]{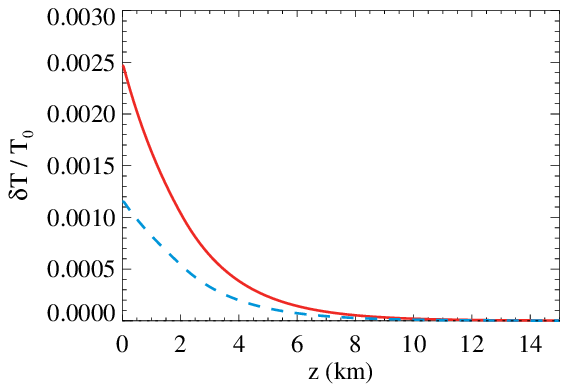}
        \includegraphics[width=0.32\hsize]{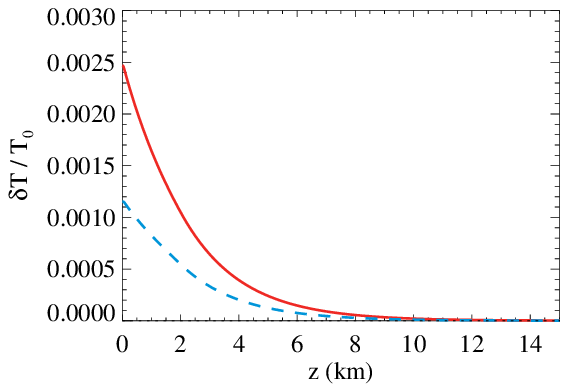}
        \caption{Same as Figure \ref{fig:secondorder_strong} but at the time $t = 10 \tau$ for a case with $\nu_{\rm{nc}} = 0.01 \omega$.}
        \label{fig:secondorder_weak}
    \end{figure*}

    The right-hand sides of Eqs. (\ref{eq:vcz_ponder}) and (\ref{eq:vnz_ponder}) represent the coupling between the first-order and the second-order perturbations. If we assume that there is a perfect coupling between the two fluids, so the collision frequency $\nu_{\rm{nc}}$ tends to infinity, Eqs. (\ref{eq:vcz_ponder}) and (\ref{eq:vnz_ponder}) simplify to
    \begin{equation} \label{eq:vpar_ideal}
        \Big[\partial_{t}^{2} - c_{\rm{eff}}^{2} \partial_{z}^{2}\Big]V_{\rm{s\parallel}} = -\frac{\partial_{t}\partial_{z} \mathcal{M}}{\rho_{\rm{c0}} \left(1 + \chi \right)}.
    \end{equation}
    If we set $\chi = 0$, we recover the equation already derived by \citet{Hollweg1971JGR....76.5155H}, \citet{Tikhonchuk10.1063/1.870975}, \citet{Terradas2004ApJ...610..523T} and \citet{Zheng2016PhyS...91a5601Z} for the case of fully ionized plasmas, in which the right-hand side represents the ponderomotive force that generates the second-order perturbations.

    The simulations represented in Figure \ref{fig:secondorder_strong} correspond to a scenario in which there is a strong but not perfect coupling, so $\nu_{\rm{nc}}$ remains finite. To obtain the appropriate differential equations, we proceed by neglecting all the terms in Eqs. (\ref{eq:vcz_ponder}) and (\ref{eq:vnz_ponder}) that do not depend on the collision frequency. Thus, we obtain the following expression:
    \begin{equation} \label{eq:vpar_strong}
        \Big[\partial_{t}^{2} - c_{\rm{eff}}^{2} \partial_{z}^{2}\Big]V_{\rm{c\parallel}} = -\frac{\partial_{t}\partial_{z} \mathcal{M} + \left(\gamma - 1 \right)\partial_{z} \mathcal{H}}{\rho_{\rm{c0}} \left(1 + \chi \right)},
    \end{equation}
    where, in comparison with Eq. (\ref{eq:vpar_ideal}), an additional term related to the energy dissipation appears. This expression has the same form as Eq. (36) from \citet{McLaughlin2011A&A...527A.149M} for the case of fully ionized plasmas, in which the dissipation is produced by resistivity, and Eq. (5.7) from \citet{Ballester2024RSPTA.38230222B} for the case of single-fluid partially ionized plasmas, in which the dissipation is produced by ambipolar diffusion.
    
    The analyses performed by \citet{McLaughlin2011A&A...527A.149M} and \citet{Ballester2024RSPTA.38230222B} show that the nonlinear coupling terms contain a combination of sinusoidal functions with twice the frequency, wavenumber, and damping rate of the first-order wave, and a non-oscillatory but exponentially decaying term, which is the cause of the bulk flow in the longitudinal direction. Here, a similar analysis allows us to explain the differences between the linearly polarized and the circularly polarized modes found in Figure \ref{fig:secondorder_strong}. First, we note that by combining Eqs. (\ref{eq:vnvc}), (\ref{eq:b1vc}), (\ref{eq:deltat2_vd}), and (\ref{eq:heating}), the heating rate $\mathcal{H}$ can be written as a function of the magnetic energy density $\mathcal{M}$. Then, by inserting Eq. (\ref{eq:by_alfven}) into the definitions from Eq. (\ref{eq:energies}), we have that the magnetic energy density of the Alfvén waves can be expressed as
    \begin{eqnarray} 
        \mathcal{M} = \frac{A_{\rm{B}}^{2}}{\mu_{0}} \exp \left(-2 k_{\rm{I}}z \right) \sin^{2} \left(k_{\rm{R}} z - \omega t + \phi_{\rm{B}} \right) \nonumber \\
        = \frac{A_{\rm{B}}^{2}}{\mu_{\rm{0}}}\exp \left(-2 k_{\rm{I}}z \right) \left[\frac{1-\cos \left(2 k_{\rm{R}} z - 2 \omega t + 2 \phi_{\rm{B}} \right)}{2} \right],
        \label{eq:mag_alfven}
    \end{eqnarray}
    which shows the dependence on twice the frequency, wavenumber, and damping rate of the first-order transverse wave. If we compute the spatial and temporal derivatives of Eq. (\ref{eq:mag_alfven}) to obtain the source terms for Eq. (\ref{eq:vpar_strong}), we get the combination of oscillatory and bulk flow terms previously mentioned.

    However, if we apply the same procedure to the case of ion-cyclotron or whistler modes, using Eqs. (\ref{eq:bx_circ}) and (\ref{eq:by_circ}), we find that
    \begin{equation} \label{eq:mag_circ}
        \mathcal{M} = \frac{A_{\rm{B}}^{2}}{2 \mu_{\rm{0}}} \exp \left(-2 k_{\rm{I}}z \right),
    \end{equation}
    which does not contain any oscillatory term and only the source of the bulk flow is retained, in agreement with the numerical results displayed on the centre and right columns of Figure \ref{fig:secondorder_strong}.

    Now, we pay attention to the case with a weaker collisional coupling, with $\nu_{\rm{nc}} = 0.01 \omega$, represented in Figure \ref{fig:secondorder_weak}. Here, the behavior of the second-order perturbations is completely dominated by the bulk flow, with no apparent oscillatory motion. This is a consequence of the strong wave damping produced by the ion-neutral interaction, which in this frequency range has a very strong impact on the nonlinearly generated wave that propagates at the sound speed. As discussed in Section IV of \citet{MartinezGomez2025PhPl...32k2101M}, in the limit of weak coupling, propagating waves have damping rates of the form $k_{\rm{I}} \sim \nu_{\rm{nc}}/(2 \mathcal{C}_{\rm{ph}})$, where $\mathcal{C}_{\rm{ph}}$ is the phase speed. Thus, waves with smaller phase speeds have larger damping rates (and, consequently, smaller damping lengths). Applying this result to the present study we would have that the nonlinearly generated wave propagating at the sound speed would be more strongly damped by collisions than the wave propagating at the Alfvén speed.

    Furthermore, Figure \ref{fig:secondorder_weak} shows that the amplitudes of the second-order perturbations of the charged fluid (red solid lines) are generally larger than those of the neutral fluid (blue dashed lines). This fact can be qualitatively understood by examining again Eqs. (\ref{eq:vcz_ponder}) and (\ref{eq:vnz_ponder}), and considering the limit of very small collision frequencies. If we neglect the terms that depend on $\nu_{\rm{nc}} \partial_{t}$ in comparison to those that depend on $\partial_{t}^{2}$, we get that the differential equations for the longitudinal velocities simplify to
    \begin{equation} \label{eq:vcpar_weak}
        \left[\partial_{t}^{2} - c_{\rm{c}}^{2} \partial_{z}^{2} \right] V_{\rm{c}\parallel} = -\frac{1}{\rho_{\rm{c}0}} \left[\frac{\left(\gamma - 1 \right)m_{\rm{n}}}{m_{\rm{n}} + m_{\rm{c}}} \partial_{z} \mathcal{H} + \partial_{t} \partial_{z} \mathcal{M}\right]
    \end{equation}
    and
    \begin{equation} \label{eq:vnpar_weak}
        \left[\partial_{t}^{2} - c_{\rm{n}}^{2} \partial_{z}^{2} \right] V_{\rm{n}\parallel} = -\frac{\left(\gamma - 1 \right)m_{\rm{c}}}{\rho_{\rm{n}0} \left(m_{\rm{n}} + m_{\rm{c}} \right)} \partial_{z} \mathcal{H},
    \end{equation}
    respectively. Now the longitudinal velocity of the neutral fluid is only affected by the heating term, while the longitudinal velocity of the charged fluid is also affected by the ponderomotive term associated to the magnetic energy density. In addition, the right-hand side of Eq. (\ref{eq:vcpar_weak}) is inversely proportional to $\rho_{\rm{c0}}$ while the right-hand side of Eq. (\ref{eq:vnpar_weak}) is inversely proportional to $\rho_{\rm{n0}}$. In weakly ionized plasma we have that $\rho_{\rm{n0}} \gg \rho_{\rm{c0}}$, so the source term for the second-order perturbations would be much larger for the charged fluid than for the neutral fluid, leading, for instance, to larger temperature increases in the charged fluid, as shown in the bottom panels of Figure \ref{fig:secondorder_weak}. 

    However, Eqs. (\ref{eq:vcpar_weak}) and (\ref{eq:vnpar_weak}) are not strictly applicable here. To obtain these approximations, we have neglected the terms with $\nu_{\rm{nc}} \partial_{t}$ and $\chi \nu_{\rm{nc}}\partial_{t}$, but since $\chi \gg 1$, the terms with $\chi \nu_{\rm{nc}} \partial_{t}$ are actually not much smaller than the terms with $\partial_{t}^{2}$. Moreover, to derive Eqs. (\ref{eq:vcz_ponder}) and (\ref{eq:vnz_ponder}) we assumed that $\Delta T\sut \approx 0$, but this condition is not fulfilled in the case with weak coupling, as shown in the bottom panels of Figure \ref{fig:secondorder_weak}. To obtain more accurate equations, the difference in temperature has to be taken into account and a sixth-order differential equation would be obtained from Eqs. (\ref{eq:rho_s_2}) - (\ref{eq:presc_2}). The derivation and analysis of that equation is out of the scope of the present study and is left for future works.

    Finally, as in Section \ref{sec:eigenfunction}, we find again that in this case there are very little differences in the results for the Alfvén, ion-cyclotron, and whistler modes. The amplitudes of the perturbations are similar for the three polarization states and the main differences can be associated with the spatial distribution of the wave energy, which is related to the phase shifts between the $x$ and $y$ components of the first-order perturbations shown in Figure \ref{fig:firstorder_weak}.
    
\section{Summary and discussion} \label{sec:concl}
    In this work, we have used a two-fluid model \citep[see, e.g.,][]{Zaqarashvili2011A&A...529A..82Z,Soler2013ApJ...767..171S,Khomenko2014PhPl...21i2901K} to describe how the linear and nonlinear evolution of Alfvénic waves in weakly ionized plasmas is modified by Hall's current and elastic collisions between ions and neutrals. We have shown that as the wave frequency approaches the Hall frequency \citep{Amagishi1993PhRvL..71..360A}, the damping rates, wave energy distribution, and heating rates strongly depend on the polarization state of the waves. For instance, the ion-cyclotron modes are more strongly damped by ion-neutral collisions than the Alfvén and whistler modes, which is a consequence of their total wave energy having a larger fraction of kinetic energy than magnetic energy \citep{Campos1992_10.1063/1.860136,MartinezGomez2025ApJ...982....4M}. In addition, the circularly polarized eigenmodes do not generate the oscillatory second-order perturbations in density, pressure and longitudinal velocity usually caused by the ponderomotive force in linearly polarized eigenmodes \citep{Hollweg1971JGR....76.5155H,Rankin1994JGR....9921291R}, but they still generate bulk flows associated with the wave energy dissipation \citep{McLaughlin2011A&A...527A.149M,Threlfall2012PhDT.......253T,Zheng2016PhyS...91c5601Z,MartinezGomez2018ApJ...856...16M,Ballester2024RSPTA.38230222B}.

    Since we have considered the cases of strong collisional coupling as well as weak coupling, the obtained results may be relevant for a wide range of astrophysical environments. For instance, the strong coupling scenario can be applied to the lower layers of the solar and stellar atmospheres, where the large densities and low ionization degrees enhance the effect of Hall's current \citep[see, e.g.,][]{Pandey2008MNRAS.385.2269P,Pandey2015MNRAS.447.3604P}. The case with a weak coupling may be of particular interest for the study of prominence to corona transition regions, where large ion-neutral drift velocities have been measured \citep[see, e.g.,][]{Khomenko2016ApJ...823..132K,Gonzalez2024A&A...681A.114G}, or the upper layers of the solar chromosphere, where the decrease in the density and collision frequencies results in a significant decoupling of neutral helium and an enhancement of the wave damping \citep{Zaqarashvili2011A&A...534A..93Z,MartinezGomez2017ApJ...837...80M,Wargnier2023ApJ...946..115W,Soler2026A&A...708A..68S,Zhang2026arXiv260105321Z}.

    Moreover, we have derived expressions that show how in weak coupling conditions, the nonlinearly generated perturbations for the neutral species are only directly caused by the energy dissipation terms, while those for the ionized species also depend on the ponderomotive force due to the magnetic pressure gradients. In addition, Eqs. (\ref{eq:vcpar_weak}) and (\ref{eq:vnpar_weak}) show that the amplitude of the longitudinal flow depends inversely on the density of the respective fluid. These approximate expressions can be used to qualitatively understand the FIP effect, in which the abundances of low-FIP elements (typically ionized) are enhanced in the upper layers of the solar atmosphere due to the action of nonlinear Alfvén waves but the abundances of high-FIP elements (typically neutral) are not affected in the same way \citep{Laming2004ApJ...614.1063L,Laming2015LRSP...12....2L,Murabito2024PhRvL.132u5201M}. However, more detailed multi-fluid simulations \citep[see, e.g.,][]{MartinezSykora2023ApJ...949..112M,MartinezSykora2026arXiv260411647M} are required to accurately describe this phenomenon.

    Lastly, we note that here we have focused only on the first-order and second-order effects of small-amplitude transverse waves, which has allowed us to study processes such as the plasma heating and the ponderomotive coupling to longitudinal motions. However, we have not considered higher-order effects such as the formation of shocks, which appear for larger amplitudes and which also play an important role in the dynamics of partially ionized plasmas \citep[see, e.g.,][]{Arber2016ApJ...817...94A,Hillier2016A&A...591A.112H,Ballester2020A&A...641A..48B,Snow2021A&A...645A..81S}. Furthermore, we have performed 1D studies of the particular case of propagation along the longitudinal direction to the background magnetic field. Therefore, it would be interesting to extend this work to the 2D and 3D scenarios with an arbitrary direction of propagation, in order to investigate the nonlinear dynamics of magnetoacoustic waves in partially ionized plasmas \citep[see, e.g.,][]{Maneva2017ApJ...836..197M,Popescu2021A&A...653A.131P,Niedziela2024A&A...691A.254N}. Another important step forward would be to include the effects of ionization and recombination in the two-fluid plasma model \citep[see, e.g.,][]{Meier2012PhPl...19g2508M,Leake2012ApJ...760..109L,Snow2021A&A...645A..81S}, which have been shown to provide an additional damping mechanism for Alfvén and slow magnetoacoustic waves \citep{Ballai2019FrASS...6...39B}, and to drive parametric resonances of Alfvén waves in weakly ionized plasmas \citep{Ballai2024RSPTA.38230226B}. These processes are also fundamental to obtain a realistic ionization degree of the plasma background considered for the study of propagation of MHD waves \citep[see, e.g.,][]{Murawski2022Ap&SS.367..111M,Kraskiewicz2025A&A...698A..74K}.
    
    
\begin{acknowledgements}
    D.M. acknowledges financial support from \textit{Govern de les Illes Balears} and \textit{Fons Social Europeu Plus (FSE+)} through the grant PD/014/2023 of the \textit{Vicenç Mut} program. This publication is part of the R+D+i project PID2023-147708NB-I00, funded by MCIN/AEI/10.13039/501100011033 and by FEDER, EU. The author also acknowledges Roberto Soler for a useful discussion on the results and the anonymous referee for helpful remarks.
\end{acknowledgements}

\bibliographystyle{aasjournal}
\bibliography{main}

\end{document}